\documentclass[aps,amsfonts,nofootinbib,preprintnumbers,nobalancelastpage]{revtex4}

\usepackage{xcolor}

\newcommand{\UH}{\mathbf{U}}
\newcommand{\DLR}{\mathbf{D}}
\begin{document}
\def\wng{{{\cal W}}_0^\gamma}
\def\wnz{{{\cal W}}_0^Z}
\def\wcg{{{\cal W}}_c^\gamma}
\def\wcz{{{\cal W}}_c^Z}
\def\wuz{{{\cal W}}_1^Z}
\def\wdz{{{\cal W}}_2^Z}
\def\wtz{{{\cal W}}_3^Z}
\def\zng{{{\cal Z}}_0^\gamma}
\def\zcg{{{\cal Z}}_c^\gamma}
\def\znz{{{\cal Z}}_0^Z}
\def\zcz{{{\cal Z}}_c^Z}

\def\Tr{\hbox{Tr}}

\preprint{\hbox{YITP-SB-16-09}}
\title{Mapping the genuine bosonic quartic couplings }

\author{O.\ J.\ P.\ \'Eboli}
\email{eboli@fma.if.usp.br}
\affiliation{Instituto de F\'{\i}sica,
             Universidade de S\~ao Paulo, S\~ao Paulo -- SP, Brazil.}

\author{M.\ C.\ Gonzalez--Garcia} \email{concha@insti.physics.sunysb.edu}
\affiliation{%
  Instituci\'o Catalana de Recerca i Estudis Avan\c{c}ats (ICREA),}
\affiliation {Departament d'Estructura i Constituents de la Mat\`eria, Universitat
  de Barcelona, 647 Diagonal, E-08028 Barcelona, Spain}
\affiliation{%
  C.N.~Yang Institute for Theoretical Physics, SUNY at Stony Brook,
  Stony Brook, NY 11794-3840, USA}

\begin{abstract}

  The larger center--of--mass energy of the Large Hadron Collider Run
  2 opens up the possibility of a more detailed study of the quartic
  vertices of the electroweak gauge bosons.  Our goal in this work is
  to classify all operators possessing quartic interactions among the
  electroweak gauge bosons that do not exhibit triple gauge--boson
  vertices associated to them. We obtain all relevant operators in the
  non-linear and linear realizations of the $SU(2)_L \otimes U(1)_Y$
  gauge symmetry.

\end{abstract}

\maketitle

\section{Introduction}

The recent discovery of a Higgs-like boson by the ATLAS and CMS
collaborations~\cite{discovery} brings us a step closer to a full
check of the standard model (SM). The SM has been subject to a large
number of precision tests during the past decades~\cite{testSM}
without any putative indication of deviations from its predictions for
the particle couplings, which in the case of fermion-gauge
interactions have been tested to close to per mil accuracy.  Since the
SM is a gauge theory based on the $SU(2)_L \otimes U(1)_Y$ group it
fixes completely the structure of the trilinear (TGC) and quartic
(QGC) electroweak gauge--boson couplings. Therefore, it is important
to establish whether these couplings indeed are in agreement with the
SM predictions.

Precise knowledge of the gauge--boson self--interactions not only can
serve to further establish the SM in the case of agreement with its
predictions, but also any observed deviation can indicate the
existence of new physics.
For instance, new heavy bosons can generate a tree level contribution
to four gauge--boson couplings while its effect in the triple--gauge
vertex would only appear at one one--loop~\cite{Arzt:1994gp}, and
consequently be suppressed with respect to the quartic one.
Moreover, the comparison of deviations in TGC and
QGC~\cite{Brivio:2013pma} can be used to determine whether the
$SU(2)_L \otimes U(1)_L$ is linearly~\cite {Buchmuller:1985jz,
  Leung:1984ni, DeRujula:1991se, Hagiwara:1993ck, Hagiwara:1993qt,
  Hagiwara:1995vp, GonzalezGarcia:1999fq,
  Grzadkowski:2010es,Passarino:2012cb} or nonlinearly~\cite{
  Appelquist:1980vg, Longhitano:1980tm, Alonso:2012px} realized in low
energy effective theory of the electroweak breaking sector.

Presently the trilinear gauge-boson couplings are known to agree with
the SM within a few percent ~\cite{LEPEWWG,TevatronTGC,LHCTGC}.  On
the other hand, there are sparse direct data on anomalous QGC and, for
a long time, the most stringent bound on QGC stemmed from their
indirect effects to the $Z$ physics via their one--loop contributions
to the oblique corrections~\cite{Eboli:1993wg, Brunstein:1996fz,
  Eboli:2006wa, Eboli:2000ad}; a situation that is starting to
change~\cite{LHC:QGCvaa, LHC:QGCvvvv, VBF:CMS, LHC:QGCwaaEXC} .
The LEP collaborations directly probed $W^+W^-\gamma\gamma$ and
$ZZ\gamma\gamma$ interactions in the reactions
$e^+ e^- \to W^+ W^- \gamma$~\cite{LEP:QGC} and
$ Z\gamma\gamma$~\cite{Achard:2002iz}. At the Tevatron, the D0
collaboration studied the $W^+W^-\gamma\gamma$ vertex in diffractive
events exhibiting dielectron and missing energy~\cite{Teva:QGC}. 
At the LHC, the ATLAS and CMS collaborations studied the production of
$V\gamma\gamma$ with $V=Z$ or $W^\pm$ to constrain the
$VV\gamma\gamma$ QGC~\cite{LHC:QGCvaa}.  Moreover, the ATLAS
collaboration analyzed the $W^+ W^-$ and $ZW^\pm$ pairs via vector
boson fusion to bound the QGC among four massive electroweak vector
bosons~\cite{LHC:QGCvvvv} while the CMS studied the $Z\gamma jj$,
$W^\pm \gamma jj$ and $W^\pm W^\pm jj$ productions to probe
QGC~\cite{VBF:CMS}. In addition to these inclusive processes the CMS
collaboration also probed the $W^+W^-\gamma\gamma$ vertex through the
exclusive $\gamma\gamma \to W^+ W^-$ production~\cite{LHC:QGCwaaEXC}.

The direct study of QGC requires either the production of three gauge
bosons or the pair production of gauge bosons in vector boson
fusion~\cite{Eboli:2000ad}. Therefore, the LHC Run 2 opens the
possibility of testing systematically anomalous QGC due to the large
center--of--mass energy.  Here we focus on genuine QGC, that is, QGC
that do not have any TGC associated to them since the best bounds on
the Wilson coefficients in the latter case are obtained from direct
study of TGC. In a scenario where the $SU(2)_L \otimes U(1)_Y$ is
realized linearly the lowest order QGC are given by dimension-eight
operators~\cite{Eboli:2006wa}. On the other hand, if the gauge
symmetry is implemented nonlinearly the lowest order QGC appear at
${\cal O}(p^4)$~\cite{Alonso:2012px,Brivio:2013pma}.

In the previous phenomenological~\cite{ Eboli:1993wg,
  Brunstein:1996fz, Eboli:2006wa, Eboli:2000ad, Eboli:2003nq,
  Belanger:1992qh, Belanger:1999aw, Stirling:1999ek, Yang:2012vv,
  Wen:2014mha} and experimental~\cite{LEP:QGC, Achard:2002iz,Teva:QGC,
  LHC:QGCvaa, VBF:CMS, LHC:QGCwaaEXC, LHC:QGCvvvv} analyses of QGC
just a partial list of effective operators has been considered. Our
goal in this work is to classify all genuine QGC in the non-linear and
linear realizations of the gauge symmetry including up to two
derivatives acting on the gauge boson fields.  This will facilitate
the translation of bounds between different notations.

The outline of the paper is as follows. We start by listing the most
general set of Lorentz structures which can be involved in quartic
gauge boson vertices containing up to two derivatives acting on the
gauge fields in Sec.~\ref{sec:lorentz}. In Sec.~\ref{sec:linear} we
present the most general effective Lagrangian which generates QGC in
scenarios in which the observed Higgs-like particle is indeed a
fundamental state belonging to an $SU(2)_L$ doublet and for which the
gauge symmetry is linearly realized. In these scenarios QGC appear at
dimension-8 independently on the number of derivatives, and we find a
total of 10 independent operators and derived the relations between
the coefficients of the generated Lorentz structures that this
implies.  Section ~\ref{sec:chiral} contains the most general
effective Lagrangian which gives rise to those QGC in scenarios with a
dynamical light Higgs for which the electroweak symmetry realization
is chiral. In this case QGC with no derivatives and two derivatives
appear at ${\cal O}(p^4)$ and ${\cal O} (p^6)$ respectively and we
find the same number of independent chiral operators as Lorentz
structures.  Finally in Sec.~\ref{sec:bounds} we summarize our
conclusions. The article is complemented with a set of appendices
containing the most lengthy expressions as well as some technical
details.

\section{General Lorentz  structures}
\label{sec:lorentz}

Without loss of generality, initially we construct the possible
Lorentz structures that are invariant under $U(1)_{\rm em}$  and
which  contain four vector bosons. We group the 
vertices in terms of the number of derivatives that they contain, and
we restrict ourselves to vertices exhibiting zero or two derivatives, 
or, equivalently, to operators with mass dimension up to six.

\subsection{Vertices with zero derivatives}

Since the $U(1)_{\text em}$ invariance requires that the photon field
appears only as part of the electromagnetic field strength, 
the zero-derivative vertices do not
contain photons, {\em i.e.}  the QGC exhibit only massive electroweak
gauge bosons.  These structures are:
\begin{equation}
\begin{array}{lcl}
 {\cal Q}^{\partial=0}_{\text WW,1} = W^{+ \mu} W^-_\mu W^{+ \nu} W^-_\nu 
&,
& {\cal Q}^{\partial=0}_{\text WW,2} = W^{+ \mu} W^{-\nu} W^+_\mu W^-_\nu  \;,
\\
 {\cal Q}^{\partial=0}_{\text WZ,1} = W^{+ \mu} W^-_\mu Z^{\nu} Z_\nu 
&,
& {\cal Q}^{\partial=0}_{\text WZ,2} = W^{+ \mu} W^{-\nu} Z_\mu Z_\nu \;,
\\
{\cal Q}^{\partial=0}_{\text ZZ} = Z^\mu Z_\mu Z^\nu Z_\nu \;.
&
\end{array}
\label{eq:0deriv}
\end{equation}
The first four structures in Eq.~(\ref{eq:0deriv}) modify the SM
quartic couplings $W^+W^-W^+W^-$ and $W^+W^-ZZ$, while the last one
leads to a QGC not present in the SM. The effective Lagrangian
containing these five structures has the general form:
\begin{equation} {\cal L}_{Q}^{\partial=0} =
  \sum_{i=1}^2 c^{0,W W}_i {\cal Q}^{\partial=0}_{W
    W,i} + \sum_{i=1}^2 c^{0,WZ}_i {\cal Q}^{\partial=0}_{WZ,i}
+c^{0,ZZ} {\cal Q}^{\partial=0}_{ZZ} \;.
\label{eq:L0d}
\end{equation}

\subsection{Vertices with 2 derivatives} 
\label{sec:lorentz2d}

The quartic vertices containing two derivatives can be classified
according to the number of photons in the vertex as 
$U(1)_{\rm em}$ requires that for each photon field at least one 
derivative must appear.
 
\subsubsection{Vertices with 2 derivatives and two photons}

The vertices with two derivatives and containing two photons are
constructed using two photon field strengths plus two $W^\pm$ or $Z$
fields.   In this case there are only four possible Lorentz 
structures:
\begin{equation}
\begin{array}{lcl}
{\cal{Q}}^{\partial=2}_{\gamma W,1} = 
F_{\mu\nu} F^{\mu\nu} W^{+ \alpha} W^-_\alpha
& ,
& {\cal{Q}}^{\partial=2}_{\gamma W,2} =
F_{\mu\nu} F^{\mu\alpha} W^{+ \nu} W^-_\alpha 
\; ,
\\
{\cal{Q}}^{\partial=2}_{\gamma Z,1}=
F_{\mu\nu} F^{\mu\nu} Z^\alpha Z_\alpha
& , 
& {\cal{Q}}^{\partial=2}_{\gamma Z,2} =
F_{\mu\nu} F^{\mu\alpha} Z^\nu Z_\alpha \; .
\label{eq:2deriv0a}
\end{array}
\end{equation}
%

\subsubsection{Vertices containing 2 derivatives and a single photon}

In this group of QGC, one derivative appears in the photon field
strength while the other derivative can appear in the anti-symmetric
combination $V_{\mu\nu} = \partial_\mu V_\nu - \partial_\nu V_\mu$
with $V_\mu = Z_\mu$ or $W^\pm_\mu$ or in the symmetric combinations
\begin{equation}
X^\pm_{\mu\nu}=\partial_\mu W^\pm_\nu+\partial_\nu W^\pm_\mu\;, 
\;\;\;\;\;\;\;\;
Y_{\mu\nu}=\partial_\mu Z_\nu+\partial_\nu Z_\mu\;.
\end{equation}
In this case one can construct nine Lorentz structures corresponding
to anomalous $\gamma Z W^+W^-$ interactions:
\begin{equation}
\begin{array}{lcll}
{\cal{Q}}^{\partial=2}_{\gamma Z W,1} =
F_{\mu\nu} Z^{\mu\nu} W^{+ \alpha} W^-_\alpha 
&,
& {\cal{Q}}^{\partial=2}_{\gamma Z W,2}  = 
F_{\mu\nu} Z^{\mu\alpha} (W^{+ \nu} W^-_\alpha + W^{- \nu} W^+_\alpha)
& , 
\\
{\cal{Q}}^{\partial=2}_{\gamma Z W,3} =
F^{\mu\nu}(W^+_{\mu\nu} W^-_\alpha Z^\alpha +  
W^-_{\mu\nu} W^+_\alpha Z^\alpha)  
&, 
& {\cal{Q}}^{\partial=2}_{\gamma Z W,4} =
F^{\mu\nu}(W^+_{\mu\alpha} W^-_\nu Z^\alpha + 
W^-_{\mu\alpha} W^+_\nu Z^\alpha) 
& ,
\\
{\cal{Q}}^{\partial=2}_{\gamma Z W,5} =
F^{\mu\nu}(W^+_{\mu\alpha} W^{- \alpha} Z_\nu + 
W^-_{\mu\alpha} W^{+ \alpha} Z_\nu)  
&,
& {\cal{Q}}^{\partial=2}_{\gamma Z W,6} =
F_{\mu\nu} Y^{\mu\alpha} (W^{+ \nu} W^-_\alpha + W^{- \nu} W^+_\alpha)
&,
\\
{\cal{Q}}^{\partial=2}_{\gamma Z W,7} = 
F_{\mu\nu} Z_\alpha (X^{+\mu\alpha} W^{-\nu} + X^{-\mu\alpha} W^{+\nu} )
& ,
& {\cal{Q}}^{\partial=2}_{\gamma Z W,8} =
F_{\mu\nu} Z^\nu (X^{+\mu\alpha} W^-_\alpha + X^{-\mu\alpha} W^+_\alpha )
& ,
\\
{\cal{Q}}^{\partial=2}_{\gamma Z W,9} =
F_{\mu\nu} Z^\mu (X^{+\alpha}_{\alpha} W^{-\nu} + X^{-\alpha}_{\alpha} W^{+\nu} ) 
& ,
&& 
\end{array}
\label{eq:2derivAZWW}
\end{equation}
while there are three Lorentz structures associated to the coupling of a single
photon to three $Z$'s
\begin{equation}
\begin{array}{lcl}
{\cal{Q}}^{\partial=2}_{\gamma Z Z,1} =
F_{\mu\nu} Z^{\mu\nu} Z^\alpha Z_\alpha  \;,\;
&
{\cal{Q}}^{\partial=2}_{\gamma Z  Z,2} =
F_{\mu\nu} Z^{\mu\alpha} Z^\nu Z_\alpha  \;,\;
&{\cal{Q}}^{\partial=2}_{\gamma Z Z,3} =  
F_{\mu\nu} Y^{\mu\alpha} Z^\nu Z_\alpha  \; .
\end{array}
\label{eq:2derivAZZZ}
\end{equation}
%
\subsubsection{Vertices with 2 derivatives without photons}   

The $W^+ W^-ZZ$ quartic interactions exhibiting two derivatives are
summarized by 29 distinct Lorentz structures:
\begin{equation}
\begin{array}{lcll}
{\cal{Q}}^{\partial=2}_{WZ,1} =
W^{+\mu\nu} W^-_{\mu\nu} Z^\alpha Z_\alpha 
&,
&{\cal{Q}}^{\partial=2}_{WZ,2}  =
W^{+\mu\nu} W^-_{\mu\alpha} Z^\alpha Z_\nu 
&,
\\
{\cal{Q}}^{\partial=2}_{WZ,3} =
W^{+\mu\nu} Z_{\mu\nu} Z^\alpha W^-_\alpha + {\rm h.c.}
&,
&{\cal{Q}}^{\partial=2}_{WZ,4} =
W^{+\mu\nu} Z_{\mu\alpha} Z_\nu W^{-\alpha} + {\rm h.c.}
&,
\\
  {\cal{Q}}^{\partial=2}_{WZ,5} =
  W^{+\mu\nu} Z_{\mu\alpha} Z^\alpha W^-_\nu + {\rm h.c.}
&,
&  {\cal{Q}}^{\partial=2}_{WZ,6} =
  Z^{\mu\nu} Z_{\mu\nu} W^+_\alpha W^{-\alpha} 
&,
\\
 {\cal{Q}}^{\partial=2}_{WZ,7} =
  Z^{\mu\nu} Z_{\mu\alpha} W^+_\nu W^{-\alpha} 
&,
&  {\cal{Q}}^{\partial=2}_{WZ,8} =
  X^{+\mu\nu} X^-_{\mu\nu} Z^\alpha Z_\alpha 
&,
\\
  {\cal{Q}}^{\partial=2}_{WZ,9} =
  X^{+\mu\nu} X^-_{\mu\alpha} Z^\alpha Z_\nu 
&,
&  {\cal{Q}}^{\partial=2}_{WZ,10} =
  X^{+\mu}_\mu X^{-\nu}_{\nu} Z^\alpha Z_\alpha 
&,
\\
  {\cal{Q}}^{\partial=2}_{WZ,11} =
 X^{+\mu}_\mu X^{-\nu\alpha} Z_\alpha Z_\nu + {\rm h.c.}
&,
&  {\cal{Q}}^{\partial=2}_{WZ,12} =
  X^{+\mu\nu} Y_{\mu\nu} Z^\alpha W^-_\alpha + {\rm h.c.}
&,
\\
  {\cal{Q}}^{\partial=2}_{WZ,13} =
  X^{+\mu\nu} Y_{\mu\alpha} Z^\alpha W^-_\nu + {\rm h.c.}
&,
&  {\cal{Q}}^{\partial=2}_{WZ,14} =
  X^{+\mu\nu} Y_{\mu\alpha} Z_\nu W^{-\alpha} + {\rm h.c.}
&,
   \\
  {\cal{Q}}^{\partial=2}_{WZ,15} =
  X^{+\mu}_\mu Y^\nu_\nu  Z_\alpha W^{-\alpha} + {\rm h.c.}
&,
&  {\cal{Q}}^{\partial=2}_{WZ,16} =
  X^{+\mu}_\mu Y^{\nu\alpha}  Z_\nu W^{-}_{\alpha} + {\rm h.c.}
&,
\\
  {\cal{Q}}^{\partial=2}_{WZ,17} =
  X^+_{\mu\alpha} Y^\nu_\nu  Z^\mu W^{-\alpha} + {\rm h.c.}
&,
&  {\cal{Q}}^{\partial=2}_{WZ,18} =
  Y^{\mu\nu} Y_{\mu\nu} W^+_\alpha W^{-\alpha} 
&,
\\
 {\cal{Q}}^{\partial=2}_{WZ,19} =
 Y^{\mu\nu} Y_{\mu\alpha} W^+_\nu W^{-\alpha} 
&,
&  {\cal{Q}}^{\partial=2}_{WZ,20} =
  Y^{\mu}_\mu Y^\nu_\nu W^+_\alpha W^{-\alpha} 
&,
\\
  {\cal{Q}}^{\partial=2}_{WZ,21} =
  Y^{\mu}_\mu Y^{\nu\alpha} W^+_\nu W^-_{\alpha} 
&,
&  {\cal{Q}}^{\partial=2}_{WZ,22} =
  W^{+\mu\nu} X^-_{\mu\alpha} Z^\alpha Z_\nu + {\rm h.c.}
&,
  \\
  {\cal{Q}}^{\partial=2}_{WZ,23} =
  W^{+\mu\nu} Y_{\mu\alpha} Z_\nu W^{-\alpha} + {\rm h.c.}
&,
&  {\cal{Q}}^{\partial=2}_{WZ,24} =
  W^{+\mu\nu} Y_{\mu\alpha} Z^\alpha W^-_\nu + {\rm h.c.}
&,
\\
  {\cal{Q}}^{\partial=2}_{WZ,25} =
  W^{+\nu\alpha} Y^\mu_\mu  Z_\nu W^-_{\alpha} + {\rm h.c.}
&,
&  {\cal{Q}}^{\partial=2}_{WZ,26} =
  X^{+\mu\nu} Z_{\mu\alpha} Z_\nu W^{-\alpha} + {\rm h.c.}
&,
\\
  {\cal{Q}}^{\partial=2}_{WZ,27} =
  X^{+\mu\nu} Z_{\mu\alpha} Z^\alpha W^-_\nu + {\rm h.c.}
&,
&  {\cal{Q}}^{\partial=2}_{WZ,28} =
  X^{+\mu}_\mu Z_{\nu\alpha}  Z^\nu W^{-\alpha} + {\rm h.c.}
&,
\\
{\cal{Q}}^{\partial=2}_{WZ,29} =
  Y^{\mu\nu} Z_{\mu\alpha} W^+_\nu W^{-\alpha} + {\rm h.c.} 
&,
& &
\end{array}
\label{eq:2deriWWZZ}
\end{equation}
where ${\rm h.c.}$ stands for the hermitian conjugate. Correspondingly,
there are 18 $W^+W^-W^+W^-$ effective vertices containing two
derivatives that are given by:
\begin{equation}
\begin{array}{lcll}
{\cal{Q}}^{\partial=2}_{WW,1} =
W^{+\mu\nu} W^-_{\mu\nu} W^{+\alpha} W^-_\alpha 
&,
& {\cal{Q}}^{\partial=2}_{WW,2} =
W^{+\mu\nu} W^-_{\mu\alpha} W^{+\alpha} W^-_\nu 
&,
\\
{\cal{Q}}^{\partial=2}_{WW,3} =
W^{+\mu\nu} W^-_{\mu\alpha} W^+_\nu W^{-\alpha} 
&,
&{\cal{Q}}^{\partial=2}_{WW,4} =
W^{+\mu\nu} W^+_{\mu\nu} W^{-\alpha} W^-_\alpha + {\rm h.c.}
&,
\\
{\cal{Q}}^{\partial=2}_{WW,5} =
W^{+\mu\nu} W^+_{\mu\alpha} W^{-\alpha} W^-_\nu + {\rm h.c.}
&,
&{\cal{Q}}^{\partial=2}_{WW,6} =
X^{+\mu\nu} X^-_{\mu\nu} W^{+\alpha} W^-_\alpha 
&,
\\
{\cal{Q}}^{\partial=2}_{WW,7} =
X^{+\mu\nu} X^-_{\mu\alpha} W^{+\alpha} W^-_\nu 
&,
&{\cal{Q}}^{\partial=2}_{WW,8} =
X^{+\mu\nu} X^-_{\mu\alpha} W^{-\alpha} W^+_\nu 
&,
\\
{\cal{Q}}^{\partial=2}_{WW,9} =
X^{+\mu}_{\mu} X^{-\nu}_\nu W^{+\alpha} W^-_\alpha 
&,
&{\cal{Q}}^{\partial=2}_{WW,10} =
X^{+\mu}_{\mu} X^{-\nu\alpha} W^+_{\nu} W^-_\alpha + {\rm h.c.}
&,
\\
{\cal{Q}}^{\partial=2}_{WW,11} =
X^{+\mu\nu} X^+_{\mu\nu} W^{-\alpha} W^-_\alpha + {\rm h.c.}
&,
&{\cal{Q}}^{\partial=2}_{WW,12} =
X^{+\mu\nu} X^+_{\mu\alpha} W^{-\alpha} W^-_\nu + {\rm h.c.}
&,
\\
{\cal{Q}}^{\partial=2}_{WW,13} =
X^{+\mu}_{\mu} X^{+\nu}_\nu W^{-\alpha} W^-_\alpha + {\rm h.c.}
&,
&{\cal{Q}}^{\partial=2}_{WW,14} =
X^{+\mu}_{\mu} X^{+\nu\alpha} W^-_{\nu} W^-_\alpha + {\rm h.c.}
&,
\\
{\cal{Q}}^{\partial=2}_{WW,15} =
W^{+\mu\nu} X^-_{\mu\alpha} W^{+\alpha} W^-_\nu + {\rm h.c.}
&,
&{\cal{Q}}^{\partial=2}_{WW,16} =
W^{+\mu\nu} X^-_{\mu\alpha} W^{-\alpha} W^+_\nu + {\rm h.c.}
&,
\\
{\cal{Q}}^{\partial=2}_{WW,17} =
X^{+\mu}_{\mu} W^-_{\nu\alpha} W^{+\nu} W^{-\alpha} + {\rm h.c.}
&,
&{\cal{Q}}^{\partial=2}_{WW,18} =
X^{+\mu\nu} W^+_{\mu\alpha} W^{-\alpha} W^-_\nu + {\rm h.c.}
&.
%
%
%
\end{array}
\label{eq:2deriWWWW}
\end{equation}
%
Finally, 7 different Lorentz structures are required to describe $ZZZZ$
vertices with two derivatives:
\begin{equation}
\begin{array}{lcll}
{\cal{Q}}^{\partial=2}_{ZZ,1} =
Z^{\mu\nu} Z_{\mu\nu} Z^{\alpha} Z_\alpha 
&,
&{\cal{Q}}^{\partial=2}_{ZZ,2} 
Z^{\mu\nu} Z_{\mu\alpha} Z^{\alpha} Z_\nu 
&,
\\
{\cal{Q}}^{\partial=2}_{ZZ,3} =
Y^{\mu\nu} Y_{\mu\nu} Z^{\alpha} Z_\alpha 
&,
& {\cal{Q}}^{\partial=2}_{ZZ,4} =
Y^{\mu\nu} Y_{\mu\alpha} Z^{\alpha} Z_\nu 
&,
\\
{\cal{Q}}^{\partial=2}_{ZZ,5} =
Y^{\mu}_{\mu} Y^\nu_{\nu} Z^{\alpha} Z_\alpha 
&,
&{\cal{Q}}^{\partial=2}_{ZZ,6} =
Y^{\mu}_{\mu} Y^{\nu\alpha} Z_{\alpha} Z_\nu 
&,
\\
{\cal{Q}}^{\partial=2}_{ZZ,7} =
Z^{\mu\nu} Y_{\mu\alpha} Z^{\alpha} Z_\nu 
&.
& &
\end{array}
\label{eq:2deriZZZZ}
\end{equation}

Altogether any effective Lagrangian possessing two derivatives and any 
four gauge bosons can be written as a combination of the seventy 
Lorentz structures above as:
\begin{eqnarray} {\cal L}^{\partial=2}_{Q} &=& 
\sum_{i=1}^2 c^{2,\gamma W}_i {\cal{Q}}^{\partial=2}_{\gamma W,i} 
+ \sum_{i=1}^2 c^{2,\gamma Z}_i {\cal{Q}}^{\partial=2}_{
    {\gamma Z},i}+
\sum_{i=1}^9 c_i^{2,\gamma Z W} {\cal{Q}}^{\partial=2}_{\gamma ZW,i}
+ \sum_{i=1}^3 c_i^{2,\gamma Z Z} {\cal{Q}}^{\partial=2}_{\gamma ZZ,i}
\nonumber \\
&&+\sum_{i=1}^{29}
  c_i^{2,WZ} {\cal{Q}}^{\partial=2}_{WZ,i} + \sum_{i=1}^{18} c_i^{2,WW}
  {\cal{Q}}^{\partial=2}_{WW,i} + \sum_{i=1}^{7} c_i^{2,ZZ} {\cal
    O}^{\partial=2}_{ZZ,i} \;.
\label{eq:Ld2}
\end{eqnarray}

It is interesting to notice that the quartic vertices in
Eqs.~(\ref{eq:2derivAZWW}), (\ref{eq:2derivAZZZ}),
(\ref{eq:2deriWWZZ}), (\ref{eq:2deriWWWW}), and (\ref{eq:2deriZZZZ})
containing $X^\pm_{\mu\nu}$ or $Y_{\mu\nu}$ have not been considered
before in the literature.

In summary, we have found five Lorentz structures without derivatives
and seventy with two derivatives.  Next we are going to build the
lowest dimension electroweak gauge invariant Lagrangian which can lead
to genuine quartic gauge boson vertices and map their coefficients to
those of these Lorentz structures.

\section {Genuine QGC in Models with an Elementary Higgs: The Linear
  Lagrangian}
\label{sec:linear}

Assuming that the new state observed in 2012 is indeed the SM Higgs
boson and belongs to a light electroweak scalar doublet, we can
construct an effective theory where the $SU(2)_L \otimes U(1)_Y$ gauge
symmetry is linearly realized~\cite{Buchmuller:1985jz, Leung:1984ni,
  DeRujula:1991se, Hagiwara:1993ck, Hagiwara:1993qt, Hagiwara:1995vp,
  GonzalezGarcia:1999fq, Grzadkowski:2010es, Passarino:2012cb} that
can be expressed as
\begin{equation}
{\cal L}_{\rm eff} = {\cal L}_{\text SM} + \sum_{n=5}^\infty
\frac{f_n}{\Lambda^{n-4}} {\cal O}_n \;\; ,
\label{l:eff}
\end{equation}
where the dimension--n operators ${\cal O}_n$ involve SM fields with
couplings $f_n$ and where $\Lambda$ is a characteristic scale.

The basic blocks for constructing the effective Lagrangian leading to
genuine QGC with the gauge symmetry realized linearly and their
transformations are:
\begin{eqnarray}
     \Phi \;, \; \hbox{ that transforms as } && \Phi^\prime = U \Phi
\\
     D_\mu \Phi \;, \; \hbox{ that transforms as }  &&
 D^\prime_\mu \Phi^\prime = U D_\mu \Phi
\\
 \widehat{W}_{\mu\nu} \equiv \sum_j W^j_{\mu\nu} \frac{\sigma^j}{2}
    \;, \; \hbox{ that transforms as }  &&
 \widehat{W}_{\mu\nu}^\prime = U  \widehat{W}_{\mu\nu} U^\dagger
\\
B_{\mu\nu} \;, \; \hbox{ that transforms as } && B_{\mu\nu}^\prime =
B_{\mu\nu}
\end{eqnarray}
where $\Phi$ stands for the Higgs doublet, $W^i_{\mu\nu}$ is the
$SU(2)_L$ field strength and $B_{\mu\nu}$ is the $U(1)_Y$ one.  Here
we denote an arbitrary gauge transformation by $U$. According to our
conventions the covariant derivative is given by
$D_\mu \Phi = (\partial_\mu + i g W^j_\mu \frac{\sigma^j}{2} + i
g^\prime B_\mu \frac{1}{2}) \Phi$
and $\sigma^j$ stand for the Pauli matrices.

Notice that the covariant derivative of $\Phi$, as well as, the field
strength tensors contain terms with at least one weak gauge boson when
we substitute $\Phi$ by its vacuum expectation value, $v$.  Therefore
the lowest dimension operators that leads to genuine quartic
interactions are dimension eight\footnote{At dimension six, quartic
  vertices are generated but they are always accompanied by  triple
  gauge boson vertices with related coefficients.}.  They can be
classified in three groups:
\begin{itemize}
\item terms that contain four covariant derivatives of the Higgs
  field; 
\item terms exhibiting two Higgs covariant derivatives and 
two field strength tensors;
\item terms presenting four field strength tensors.
\end{itemize}
Here we focus on operators containing up to two derivatives,
therefore, we will not analyze the last class, however, we present
these operators in the Appendix~\ref{ap:4FS} for the sake of
completeness.

\subsection{ Operators containing only $D_\mu\Phi$}

There are three independent operators belong to this class\footnote{We
  are using the conventions of Ref. \cite{Eboli:2006wa}, however, the
  operator ${\cal O}_{S,2}$ has been included.  }:
\begin{equation}
\begin{array}{l}
  {\cal O}_{S,0} = 
\left [ \left ( D_\mu \Phi \right)^\dagger
 D_\nu \Phi \right ] \times 
\left [ \left ( D^\mu \Phi \right)^\dagger
D^\nu \Phi \right ]
=\frac{g^{{4}} v^4}{16} \left[
 {\cal Q}^{\partial=0}_{WW,2}
+\frac{1}{c_w^2}  {\cal Q}^{\partial=0}_{WZ,2} 
+\frac{1}{4c_w^4}  {\cal Q}^{\partial=0}_{ZZ}  \right]
\; , 
\\
  {\cal O}_{S,1} =
 \left [ \left ( D_\mu \Phi \right)^\dagger
 D^\mu \Phi  \right ] \times
\left [ \left ( D_\nu \Phi \right)^\dagger
D^\nu \Phi \right ]
= \frac{g^4 v^{{4}}}{16} \left[
 {\cal Q}^{\partial=0}_{WW,1}
+\frac{1}{c_w^2}  {\cal Q}^{\partial=0}_{WZ,1} 
+\frac{1}{4c_w^4}  {\cal Q}^{\partial=0}_{ZZ}  \right]
\; , 
\\
  {\cal O}_{S,2} =
 \left [ \left ( D_\mu \Phi \right)^\dagger
 D_\nu \Phi  \right ] \times
\left [ \left ( D^\nu \Phi \right)^\dagger
D^\mu \Phi \right ]
= \frac{g^{4} v^4}{16} \left[
 {\cal Q}^{\partial=0}_{WW,1}
+\frac{1}{c_w^2}  {\cal Q}^{\partial=0}_{WZ,2} 
+\frac{1}{4c_w^4}  {\cal Q}^{\partial=0}_{ZZ}  \right]
\; . 
\end{array}
\label{eq:dphi}
\end{equation}
The corresponding effective Lagrangian can be mapped to that in
Eq.~(\ref{eq:L0d}) with coefficients related as:
\begin{equation}
\begin{array}{llll}
c^{0,WW}_1 =\frac{g^4 v^4}{16} \left [ 
\frac{f_{S,1}}{\Lambda^4}  + \frac{f_{S,2}}{\Lambda^4} \right ] 
&,
&c^{0,WW}_2 =\frac{g^4 v^4}{16}   \frac{f_{S,0}}{\Lambda^4}
& , 
\\ 
c^{0,WZ}_1=\frac{g^4 v^4}{16 c_W^2} \frac{f_{S,1}}{\Lambda^4}  
&, 
& c^{0,WZ}_2=\frac{g^4 v^4}{16 c_w^2}  
\left[ \frac{f_{S,0}}{\Lambda^4} + \frac{f_{S,2}}{\Lambda^4} \right] 
&,
\\
 c^{0,ZZ}= \frac{1}{4  c_w^4} 
\left(  c_1^{0,WW} + c_2^{0,WW}
  \right)
&,
&&
\end{array}
\label{eq:lin1}
\end{equation}
where $s_w$ ($c_w$) stand for the sine (cosine) of the weak mixing
angle $\theta_w$ verifying $\tan\theta_w=g/g'$.  Notice that the
linear realization of the symmetry leads to correlations between the
Wilson coefficients appearing in Eq.~(\ref{eq:L0d}).

\subsection{ Operators containing $D_\mu\Phi$ and field strength} 

This class possesses seven operators\footnote{We follow the notation
  in Ref.~\cite{Eboli:2006wa} but we notice that in there an
  additional operator ${\cal O}_{M,6}$ was listed which we found to be
  redundant.}:
\begin{equation}
\begin{array}{lcll}
 {\cal O}_{M,0} =   \hbox{Tr}\left [ \widehat{W}_{\mu\nu} \widehat{W}^{\mu\nu} \right ]
\times  \left [ \left ( D_\beta \Phi \right)^\dagger
D^\beta \Phi \right ]
&,& 
 {\cal O}_{M,1} 
=   \hbox{Tr}\left [ \widehat{W}_{\mu\nu} \widehat{W}^{\nu\beta} \right ]
\times  \left [ \left ( D_\beta \Phi \right)^\dagger
D^\mu \Phi \right ]
&,
\\
 {\cal O}_{M,2} =   \left [ B_{\mu\nu} B^{\mu\nu} \right ]
\times  \left [ \left ( D_\beta \Phi \right)^\dagger
D^\beta \Phi \right ]
&,&
 {\cal O}_{M,3} =   \left [ B_{\mu\nu} B^{\nu\beta} \right ]
\times  \left [ \left ( D_\beta \Phi \right)^\dagger
D^\mu \Phi \right ]
&,
\\
  {\cal O}_{M,4} = \left [ \left ( D_\mu \Phi \right)^\dagger \widehat{W}_{\beta\nu}
 D^\mu \Phi  \right ] \times B^{\beta\nu}
&,&
  {\cal O}_{M,5} = \left [ \left ( D_\mu \Phi \right)^\dagger \widehat{W}_{\beta\nu}
 D^\nu \Phi  \right ] \times B^{\beta\mu}+ {\rm h.c.}
&,
\\
  {\cal O}_{M,7} = \left [ \left ( D_\mu \Phi \right)^\dagger \widehat{W}_{\beta\nu}
\widehat{W}^{\beta\mu} D^\nu \Phi  \right ]  
&.&&
\end{array}
\label{eq:lind2}
\end{equation}
These seven operators involve 23 of the 70 possible Lorentz structures
as explicitly given in Appendix \ref{app:linVSlor}.  In particular
none of the structures with symmetric gauge boson tensors are
generated.  The corresponding effective Lagrangian can be written in
the form of Eq.~(\ref{eq:Ld2}) where the 23 coefficients of the
Lorentz structures can be expressed in terms of 7 of them. For
example, these can be chosen to be $c^{2,WW}_{1,2}$,
$c^{2,WZ}_{2,3,6,7}$ and $c^{2,\gamma W}_{2}$ with
\begin{eqnarray}
c^{2,WW}_1&=&\frac{g^2 v^2}{8}\frac{2f_{M,0}}{\Lambda^4}\;, 
\nonumber\\
c^{2,WW}_2&=&-\frac{g^2 v^2}{8}\frac{f_{M,1}}{\Lambda^4}\; , 
\nonumber\\
c^{2,WZ}_2&=&\frac{g^2 v^2}{8}\frac{1}{c_w^2}
\frac{-f_{M,1}+\frac{1}{2}f_{M,7}}{\Lambda^4}
\;,
\nonumber \\
c^{2,WZ}_3&=& \frac{g^2 v^2}{8}\frac{s_w}{c_w} \frac{f_{M,4}}{\Lambda^4}\;,
\label{eq:lin2}\\
c^{2,WZ}_6&=& \frac{g^2 v^2}{8}\left[c_w^2\frac{f_{M,0}}{\Lambda^4}
+2 s_w^2 \frac{f_{M,2}}{\Lambda^4}
- s_w c_w \frac{f_{M,4}}{\Lambda^4}\right]\; , 
\nonumber \\
c^{2,WZ}_7&=& \frac{g^2 v^2}{8}\left[-c_w^2\frac{f_{M,1}}{\Lambda^4}
- s_w^2 \frac{f_{M,3}}{\Lambda^4}
- 2 s_w c_w \frac{f_{M,5}}{\Lambda^4}
+\frac{1}{2}c_w^2 \frac{f_{M,7}}{\Lambda^4}\right]\; , 
\nonumber \\
  c^{2,\gamma W}_2&=& \frac{g^2 v^2}{8}\left[-s_w^2\frac{f_{M,1}}{\Lambda^4}
- c_w^2 \frac{f_{M,3}}{\Lambda^4}
+2 s_w c_w \frac{f_{M,5}}{\Lambda^4}
+\frac{1}{2}s_w^2 \frac{f_{M,7}}{\Lambda^4}\right]\; , 
\nonumber 
\end{eqnarray}
and the remaining 16 coefficients satisfy the following relations
\begin{equation}
\begin{array}{llll}
c^{2,WZ}_1  =\frac{1}{2 c_w^2} c^{2,WW}_1 &,&
c^{2,\gamma Z}_1= 
\frac{1}{2c_w^2} c^{2,\gamma W}_1- c^{2,WZ}_3 &,
\\
c^{2,ZZ}_1 =  
\frac{1}{2c_w^2} c^{2,WZ}_6+ c^{2,WZ}_3 
&,&
c^{2,\gamma ZW}_3 = -\frac{c_w}{s_w} c^{2,WZ}_3 &, 
\\
c^{2,\gamma ZW}_1= 
2 s_w c_w \left[c_1^{2,WW}-c_6^{2,WZ}-c_1^{2,\gamma W}+\frac{
c_w^2-s_w^2}{2 s^2_w} c_3^{2,WZ}\right]
&,& 
c^{2,\gamma ZZ}_1= \frac{1}{2 c_w^2} c_1^{2,\gamma ZW} 
-\frac{c_w^2-s_w^2}{s_wc_w} c_3^{2,WZ} &,
\\
c_1^{2,\gamma W} = \frac{1}{2 s_w^2} \left[ 2 c_w^2(
  c_3^{2,WZ}+c_6^{2,WZ}) -  (c_w^2 -s_w^2) c_1^{2,WW}   \right]   
 &,& 
c_3^{2,WW} = 2c_w^2 c_2^{2,WZ} -c_2^{2,WW} 
&,
\\
c_2^{2,ZZ}=\frac{1}{2c_w^2}\left[(c_w^2-s_w^2)(2 c_w^2 c_2^{2,WZ}-c_7^{2,WZ})
+2 s_w^2 c_2^{2,\gamma W} \right] &,&
c_2^{2,\gamma Z} = -c_2^{2,ZZ}+\frac{1}{c_w^2}(c_7^{2,WZ}+c_2^{2,\gamma W})
&,\\
c_5^{2,WZ} = \frac{1}{2}\left[-c_2^{2,WZ}-c_7^{2,WZ}+\frac{s_w^2}{c_w^2}c_2^{2,\gamma W}+ 2 c_2^{2,WW}\right] &,& 
c_4^{2,WZ}= c_5^{2,WZ}+2 (c_w^2 c_2^{2,WZ}- c_2^{2,WW})
&,\\
c_2^{2,\gamma ZW}
=\frac{1}{2 s_w c_w} 
\left[c_w^2(c_2^{2,WZ}-c_7^{2,WZ})-s_w^2  c_2^{2,\gamma W} \right] &,&
c_2^{2,\gamma ZZ}=\frac{1}{c_w^2}c_2^{2,\gamma ZW}-\frac{c_w^2-s_w^2}{c_w s_w}
(c_5^{2,WZ}+c_4^{2,WZ} ) &,\\
c_{4,5}^{2,\gamma ZW}=-\frac{c_w}{2s_w}(c_5^{2,WZ}+c_4^{2,WZ} )\pm\frac{s_w}{c_w}
 (c_w^2 c_2^{2WZ}-c_2^{2,WW}) &.&
\end{array}
\label{eq:lin3}
\end{equation}

\section{ Genuine QGC in Models with a Dynamical Higgs: The Chiral
  Lagrangian}
\label{sec:chiral}

In dynamical Higgs scenarios, the Higgs particle is a composite field
which happens to be a pseudo-Nambu-Goldstone boson (PNG) of a global
symmetry exact at scales $\Lambda_{\text strong}$ that corresponds to
the masses of the lightest strong resonances.  Because the Higgs-like
particle is a PNG, the effective Lagrangian is non-linear or
``chiral": a derivative expansion \cite{Weinberg:1978kz,
  Feruglio:1992wf, Appelquist:1980vg, Longhitano:1980tm} with a global
$SU(2)_L \otimes SU(2)_R$ symmetry broken to the diagonal $SU(2)_c$.
The effective low-energy chiral Lagrangian is entirely written in
terms of the SM fermions and gauge bosons and of the physical Higgs
$h$.  In this scenario, the basic building block at low energies is a
dimensionless unitary matrix transforming as a bi-doublet of the
global symmetry:
\begin{equation}
\UH(x)=e^{i\sigma_a \pi^a(x)/v}\; , \qquad \qquad  \UH(x) \rightarrow L\, \UH(x) R^\dagger\;,
\end{equation}
where $L$, $R$ denote $SU(2)_{L,R}$ global transformations,
respectively and $\pi^a$ are the goldstone bosons.  Its covariant
derivative reads
\begin{equation}
\DLR_\mu \UH(x) \equiv \partial_\mu \UH(x) +ig \frac{\sigma^j}{2}
W^i_{\mu}(x)\UH(x) - \frac{ig'}{2}  B_\mu(x) \UH(x)\sigma_3 \; .
\end{equation}

We define the vector chiral field and its covariant derivative as
\begin{eqnarray}
V_\mu&\equiv&  \left(\DLR_\mu\UH\right)\UH^\dagger \;,
\\
D_\mu V_\alpha &=& \partial_\mu V_\alpha + i g [ W_\mu , V_\alpha]  \;,
\end{eqnarray}
and the scalar chiral field $T\equiv\UH\sigma_3\UH^\dag$. These three
objects transform in the adjoint of $SU(2)_L$. Moreover, the Higgs
field $h$ is a singlet under the global symmetry.

In our framework, we consider genuine QGC that appear at
${\cal O}(p^4)$ and ${\cal O}(p^6)$ and are invariant under $CP$.  The
$CP$ transformation properties of our building blocks can be easily
obtained once we know that ~\cite{Longhitano:1980tm}
\begin{equation}
\begin{array}{lclc}
CP B_\mu(\vec{x},t) (CP)^{-1} = - B_\mu(-\vec{x},t) 
&,
&CP W^i_\mu(\vec{x},t) (CP)^{-1} =  \sigma^2 W^i_\mu(-\vec{x},t) \sigma^2
&,
\\
CP T(\vec{x},t) (CP)^{-1} = - \sigma^2 T(-\vec{x},t) \sigma^2 
&,
&CP V_\mu(\vec{x},t) (CP)^{-1} =  \sigma^2 V_\mu(-\vec{x},t) \sigma^2
&. 
\end{array}
\label{eq:cp}
\end{equation}
Our choice of phases are such that $D_\mu U$ has a well-defined
transformation under $CP$. From the above equation we can learn that
the $CP$ conserving QGC are the ones exhibiting an even number of
$T$'s and $B_\mu$'s.

The building blocks that we use to construct genuine $CP$ conserving
QGC can be classified according the mass dimension of the operator
($D$)~\cite{Longhitano:1980tm}.  Here we list all building block
operators needed to construct up to ${\cal O}(p^6)$ quartic operators,
as well as their expressions in the unitary gauge. There is just one
operator with mass dimension one
\begin{equation}
\hbox{Tr} \left [ T V_\mu \right ] = i \frac{g}{c_w} Z_\mu \;.
\label{eq:TV}
\end{equation}
On the other hand, there are five $D=2$ building blocks, however, only
4 of them appear in $CP$ invariant quartic operators:
\begin{eqnarray}
B_{\mu\nu}  &=& c_w F_{\mu\nu} - s_w Z_{\mu\nu} \;,
\\
\hbox{Tr} [ T D_{\mu\nu}  ] &=& i \frac{g}{c_w}  Y_{\mu\nu} \;,
\\
\hbox{Tr} [ V_\mu V_\nu] &=& - \frac{g^2}{2} \left ( \frac{1}{c_w^2}
                            Z_\mu Z_\nu + W^+_\mu W^-_\nu + W^-_\mu
                            W^+_\nu \right ) \;,
\label{eq:VV}
\\
\hbox{Tr} [ T \widehat{W}_{\mu\nu} ] &=& c_w Z_{\mu\nu} + s_w F_{\mu\nu} \;,
\end{eqnarray}
where we define the symmetric combination
$ D_{\mu \nu} \equiv D_\mu V_\nu+D_\nu V_\mu$.

Just two of the eight $D=3$ basic operators appear in $CP$ invariant
quartic vertices:
\begin{eqnarray}
%
\hbox{Tr} [ V_\mu D_{\nu\lambda} ] &=&
- \frac{g^2}{2} \left ( \frac{1}{c^2_W} Z_\mu Y_{\nu\lambda}
+ W^+_\mu X^-_{\nu\lambda} + W^-_\mu X^+_{\nu\lambda} \right ) \;,
\\
\hbox{Tr} [ V_\mu \widehat{W}_{\nu\lambda} ] &=&
i \frac{g}{2} \left ( \frac{1}{c_w} Z_\mu (c_w Z_{\nu\lambda} + s_w
     F_{\nu\lambda} )  + W^+_\mu W^- _{\nu\lambda} + W^-_\mu
                                             W^+ _{\nu\lambda}  \right ) \;.
\end{eqnarray}
Of the eleven possible $D=4$ operators just three of them contribute to
$CP$ conserving quartic vertices:
\begin{eqnarray}
  \hbox{Tr} [ D_{\mu\nu} D_{\alpha\beta} ] &=& - \frac{g^2}{2}
\left (  \frac{1}{c_w^2} Y_{\mu\nu} Y_{\alpha\beta} 
+ X^+_{\mu\nu}  X^-_{\alpha\beta} 
+ X^-_{\mu\nu}  X^+_{\alpha\beta}    \right ) \;,
\\
  \hbox{Tr} [ \widehat{W}_{\mu\nu} \widehat{W}_{\alpha\beta} ] &=&\frac {1}{2}
\left [ (c_w Z_{\mu\nu} +s_w F_{\mu\nu})
(c_w Z_{\alpha\beta} +s_w F_{\alpha\beta})
+ W^+_{\mu\nu}  W^-_{\alpha\beta} 
+ W^-_{\mu\nu}  W^+_{\alpha\beta}    \right ] \;,
\\
  \hbox{Tr} [ \widehat{W}_{\mu\nu} D_{\alpha\beta} ] &=& \frac{i g}{2}
\left [  \frac{1}{c_w} (c_w Z_{\mu\nu} +s_w F_{\mu\nu}) Y_{\alpha\beta} 
+ W^+_{\mu\nu}  X^-_{\alpha\beta} 
+ W^-_{\mu\nu}  X^+_{\alpha\beta}    \right ] \;.
\end{eqnarray}
It is interesting to notice that no dimension-five operador can give
rise to $p^6$ $CP$-conserving QGC.

Using the $D=1,2,3,4$ building blocks, we construct all possible
$CP$-conserving operators for genuine QGC, and then we remove those which can be
related by total derivatives. For instance, the relation
\begin{equation}
\begin{array}{ll}
\partial_\nu \left \{ 
\hbox{Tr}[V_\mu D_\lambda V_\rho ] ~\hbox{Tr} [  V_\alpha V_\beta]   
\right \} =&
\hbox{Tr}[D_\nu V_\mu D_\lambda V_\rho ] ~\hbox{Tr} [  V_\alpha V_\beta]   +
\hbox{Tr}[V_\mu D_\nu  D_\lambda V_\rho ] ~\hbox{Tr} [  V_\alpha
            V_\beta]   +
\\
&
\hbox{Tr}[V_\mu D_\lambda V_\rho ] ~\hbox{Tr} [ D_\nu  V_\alpha V_\beta]   +
\hbox{Tr}[V_\mu D_\lambda V_\rho ] ~\hbox{Tr} [  V_\alpha D_\nu V_\beta]   
\end{array}
\end{equation}
can be used to eliminate operators that contain the building block
$\hbox{Tr}[V_\mu D_\nu D_\lambda V_\rho ]$.  To further reduce the
number of equivalent operators we also use the relation
\begin{equation}
D_\mu V_\nu-D_\nu V_\mu
=i g \widehat{W}_{\mu\nu}-\frac{i}{2}g'B_{\mu\nu}T 
+ [V_\mu, V_\nu] \;.
\end{equation}
Moreover, we introduce a factor $i \, g$ and $i \, g'$ in each
operator containing $\widehat W_{\mu\nu}$ and $B_{\mu\nu}$
respectively in order to have consistent global powers of coupling
constants.

\subsection{QGC at ${\cal O}(p^4)$}

The lowest order genuine quartic operators are ${\cal O}(p^4)$, and
there are two operators which respect the $SU(2)_c$ custodial
symmetry, as well as $C$ and $P$ that are given by\footnote{We follow
  the notation of Refs.~\cite{Alonso:2012px,Brivio:2013pma}. }
\begin{equation}
\begin{array}{l}
{\cal P}_6 = \Tr [ V^\mu V_\mu] \Tr[V^\nu V_\nu] {\cal F}_6(h)
= g^4\left[\frac{1}{4c_w^4}  {\cal O}^0_{ZZ}
+ {\cal O}^0_{WW,1} +\frac{1}{c_w^2}  {\cal O}^0_{WZ,1}\right] {\cal F}_6(h)\; , 
\\
{\cal P}_{11} = \Tr [ V^\mu V^\nu] \Tr[V_\mu V_\nu] {\cal F}_{11}(h)
=g^4\left[
\frac{1}{4c_w^4}  {\cal O}^0_{ZZ}
+ \frac{1}{2} {\cal O}^0_{WW,1}
+ \frac{1}{2} {\cal O}^0_{WW,2}
+ \frac{1}{c_w^2}  {\cal O}^0_{WZ,2}\right]{\cal F}_{11}(h) \;,  
\label{eq:p41}
\end{array}
\end{equation}
and 3 additional $CP$ conserving operators that violate $SU(2)_c$:
\begin{equation}
\begin{array}{l}
{\cal P}_{23} = \Tr [ V^\mu V_\mu] (\Tr[ T V_\nu])^2 {\cal F}_{23}(h)
=g^4\left[\frac{1}{2c_w^4}  {\cal O}^0_{ZZ}
+ \frac{1}{c_w^2}  {\cal O}^0_{WZ,1}\right]{\cal F}_{23}(h) \; , 
\\
{\cal P}_{24} = \Tr [ V^\mu V^\nu] \Tr[ T V_\mu] \Tr[T V_\nu]
{\cal F}_{24}(h)
=g^4\left[\frac{1}{2c_w^4}  {\cal O}^0_{ZZ}
+ \frac{1}{c_w^2}  {\cal O}^0_{WZ,2}\right]{\cal F}_{24}(h)\; , 
\\ 
{\cal P}_{26} = ( \Tr[ T V_\mu] \Tr[T V_\nu] )^2{\cal F}_{26}(h)
= \frac{g^4}{c_w^4}  {\cal O}^0_{ZZ}{\cal F}_{26}(h)
\; . 
\label{eq:p42}
\end{array}
\end{equation}
${\cal F}_i(h)$ are generic functions parametrizing the
chiral-symmetry breaking interactions of $h$ which can be expanded as
${\cal F}_i(h)\equiv 1+2\tilde a_i\frac{h}{v}+\tilde
b_i\frac{h^2}{v^2}+\ldots $.
As we are looking for operators whose lowest order vertex contains
four gauge bosons, we will be only concerned by the constant term.  So
the most general Lagrangian at ${\cal O}(p^4)$ for genuine QGC is
\begin{equation}
{\cal L}^{p=4}_Q=\sum_{i=6,11,23,24,26} c^{p=4}_i {\cal P}_i \;.
\end{equation} 
From Eqs.~(\ref{eq:p41}) and ~(\ref{eq:p42}) we see that the above
Lagrangian leads to quartic gauge couplings which do not contain
photons.  We also see that there are five operators matching five
independent Lorentz structures that do not exhibit derivatives. In
Ref.~\cite{Longhitano:1980tm} we can find the $p^4$ QGC assuming that
there is no light Higgs-like state and this corresponds to the limit
${\cal F}_i \to 1$ in our framework.  The correspondence between
between the Wilson coefficients our notation and the one of
Ref.~\cite{Longhitano:1980tm} is
\begin{equation}
\alpha_4 = c_{11}^{p=4} \;\;,\;\;
\alpha_5 = c_{6}^{p=4} \;\;,\;\;
\alpha_6 = c_{24}^{p=4} \;\;,\;\;
\alpha_7 = c_{23}^{p=4} \;\;,\;\;
\alpha_{10} = c_{26}^{p=4} \;\;.
\end{equation}

\subsection{QGC at ${\cal O}(p^6)$} 

At order ${\cal O}(p^6)$, there is the emergence of genuine QGC
containing photons as well as only four electroweak gauge bosons with
two derivatives acting on them.  As in Ref.~\cite{Longhitano:1980tm},
we construct the $p^6$ operators for QGC combining the $D=1,2,3,$ and
$4$ building blocks defined above. Without loss of generality we write
the corresponding Lagrangian as
\begin{equation}
{\cal L}^{p=6}_{\cal Q}=\sum_i c^{p=6}_i{\cal T}^{p=6}_i 
{\cal  F}^{p=6}_i(h)
\end{equation}
where ${\cal T}^{p=6}_i$ are the ${\cal O}(p^6)$ operators constructed
with the blocks defined above, and we denote by ${\cal F}^{p=6}_i(h)$
the corresponding arbitrary function parametrizing the $h$ couplings.
As already mentioned we will be only concerned with the first term of
its expansion ${\cal F}^{p=6}_i(h)=1$.

\subsubsection{${\mathbf (D=1)^2(D=2)^2}$ terms}

There are twelve independent operators in this category:
\begin{equation}
\begin{array}{lclc}
{\cal T}^{p=6}_1 = \Tr[T {\cal D}_{\mu\nu}]\Tr[T {\cal D}^{\mu\nu}]
\Tr[T V^\alpha] \Tr[T V_\alpha] 
&,
&{\cal T}^{p=6}_2 = \Tr[T {\cal D}_{\mu\nu}]\Tr[T {\cal D}^{\mu\alpha}]
\Tr[T V^\nu] \Tr[T V_\alpha] 
&,
\\
{\cal T}^{p=6}_3 = \Tr[T {\cal D}^\mu_{\mu}]\Tr[T {\cal D}^{\nu}_\nu]
\Tr[T V^\alpha] \Tr[T V_\alpha] 
&,
&{\cal T}^{p=6}_4 = \Tr[T {\cal D}^\mu_{\mu}]\Tr[T {\cal D}^{\nu\alpha}]
\Tr[T V_\nu] \Tr[T V_\alpha] 
&,
\\
{\cal T}^{p=6}_5 = -g^2\,\Tr[T \widehat{W}_{\mu\nu}]\Tr[T \widehat{W}^{\mu\nu}]
\Tr[T V^\alpha] \Tr[T V_\alpha] 
&,
&{\cal T}^{p=6}_6 = -g^2\,\Tr[T \widehat{W}_{\mu\nu}]\Tr[T \widehat{W}^{\mu\alpha}]
\Tr[T V^\nu] \Tr[T V_\alpha] 
&,
\\
{\cal T}^{p=6}_7 =  i\, g\, Tr[T \widehat {W}_{\mu\nu}]\Tr[T {\cal D}^{\mu\alpha}]
\Tr[T V^\nu] \Tr[T V_\alpha] 
&,
&{\cal T}^{p=6}_8 = -{g'}^2\,{B}_{\mu\nu} {B}^{\mu\nu}
\Tr[T V^\alpha] \Tr[T V_\alpha] 
&,
\\
{\cal T}^{p=6}_9 = -{g'}^2\, B_{\mu\nu} B^{\mu\alpha}
\Tr[T V^\nu] \Tr[T V_\alpha] 
&,
&{\cal T}^{p=6}_{10} = i\, g' \, B_{\mu\nu}\Tr[T {\cal D}^{\mu\alpha}]
\Tr[T V^\nu] \Tr[T V_\alpha] 
&,
\\
{\cal T}^{p=6}_{11} = - g g' \,B_{\mu\nu}\Tr[T \widehat{W}^{\mu\nu}]
\Tr[T V^\alpha] \Tr[T V_\alpha] 
&,
&{\cal T}^{p=6}_{12} = -g g' \,B_{\mu\nu}\Tr[T \widehat{W}^{\mu\alpha}]
\Tr[T V^\nu] \Tr[T V_\alpha] 
&.
\end{array}
\end{equation}
Notice that all the effective Lagrangian in this class violate
$SU(2)_c$ since they contain the $T$ field.  We present the relations
between these operators and the Lorentz structures in
Appendix~\ref{app:nlVSlor} that allow us to see that all the operators
in this group contain only neutral gauge bosons.

\subsubsection{${\mathbf (D=1)(D=2)(D=3)}$ terms } 

This group contains twenty one operators that violate $SU(2)_c$ due to
the presence of $T$:
\begin{equation}
\begin{array}{lclc}
{\cal T}^{p=6}_{13} = 
\Tr[T V_\alpha] \Tr[T {\cal D}_{\mu\nu}]\Tr[ V^\alpha{\cal D}^{\mu\nu}] 
&,
&{\cal T}^{p=6}_{14} = 
\Tr[T V_\alpha] \Tr[T {\cal D}_{\mu\nu}]\Tr[ V^\nu{\cal D}^{\mu\alpha}] 
&,
\\
{\cal T}^{p=6}_{15} = 
\Tr[T V^\nu] \Tr[T {\cal D}_{\mu\nu}]\Tr[ V_\alpha{\cal D}^{\mu\alpha}] 
&,
&{\cal T}^{p=6}_{16} = 
\Tr[T V_\alpha] \Tr[T {\cal D}_{\mu}^\mu]\Tr[ V^\alpha{\cal D}^{\nu}_\nu] 
&,
\\
{\cal T}^{p=6}_{17} = 
\Tr[T V_\nu] \Tr[T {\cal D}_{\mu}^\mu]\Tr[ V_\alpha{\cal D}^{\nu\alpha}] 
&,
&{\cal T}^{p=6}_{18} = 
\Tr[T V^\nu] \Tr[T {\cal D}_{\nu\alpha}]\Tr[ V^\alpha{\cal D}^{\mu}_\mu] 
&,
\\
{\cal T}^{p=6}_{19} =  i\, g \,
\Tr[T V_\alpha] \Tr[T {\cal D}_{\mu\nu}]\Tr[ V^\nu \widehat{W}^{\mu\alpha}] 
&,
&{\cal T}^{p=6}_{20} = i\, g \,
\Tr[T V^\nu] \Tr[T {\cal D}_{\mu\nu}]\Tr[ V_\alpha \widehat{W}^{\mu\alpha}] 
&,
\\
{\cal T}^{p=6}_{21} = i\, g \,
\Tr[T V_\nu] \Tr[T {\cal D}_{\mu}^\mu]\Tr[ V_\alpha \widehat{W}^{\nu\alpha}] 
&,
&{\cal T}^{p=6}_{22} = i\, g \,
\Tr[T V_\alpha] \Tr[T \widehat{W}_{\mu\nu}]\Tr[ V^\nu{\cal D}^{\mu\alpha}] 
&,
\\
{\cal T}^{p=6}_{23} = i\, g \,
\Tr[T V^\nu] \Tr[T \widehat{W}_{\mu\nu}]\Tr[ V_\alpha{\cal D}^{\mu\alpha}] 
&,
&{\cal T}^{p=6}_{24} = i\, g \,
\Tr[T V^\nu] \Tr[T \widehat{W}_{\nu\alpha}]\Tr[ V^\alpha{\cal D}^{\mu}_\mu] 
&,
\\
{\cal T}^{p=6}_{25} =  -g^2 \, 
\Tr[T V_\alpha] \Tr[T \widehat{W}_{\mu\nu}]\Tr[ V^\alpha \widehat{W}^{\mu\nu}] 
&,
&{\cal T}^{p=6}_{26} = -g^2 \, 
\Tr[T V_\alpha] \Tr[T \widehat{W}_{\mu\nu}]\Tr[ V^\nu \widehat{W}^{\mu\alpha}] 
&,
\\
{\cal T}^{p=6}_{27} = -g^2 \, 
\Tr[T V^\nu] \Tr[T \widehat{W}_{\mu\nu}]\Tr[ V_\alpha \widehat{W}^{\mu\alpha}] 
&,
&{\cal T}^{p=6}_{28} = i\, g' \,
\Tr[T V_\alpha] B_{\mu\nu}\Tr[ V^\nu{\cal D}^{\mu\alpha}] 
&,
\\
{\cal T}^{p=6}_{29} = i\, g' \,
\Tr[T V^\nu] B_{\mu\nu}\Tr[ V_{\alpha}{\cal D}^{\mu\alpha}] 
&,
&{\cal T}^{p=6}_{30} = i\, g' \,
\Tr[T V^\nu] B_{\nu\alpha}\Tr[ V^\alpha{\cal D}^{\mu}_\mu] 
&,
\\
{\cal T}^{p=6}_{31} = -g g' \,
\Tr[T V_\alpha] B_{\mu\nu}\Tr[ V^\alpha \widehat{W}^{\mu\nu}] 
&,
&{\cal T}^{p=6}_{32} = -g g' \,
\Tr[T V_\alpha] B_{\mu\nu}\Tr[ V^\nu \widehat{W}^{\mu\alpha}] 
&,
\\
{\cal T}^{p=6}_{33} = -g g' \,
\Tr[T V^\nu] B_{\mu\nu} \Tr[ V_\alpha \widehat{W}^{\mu\alpha}] 
&.
&&
\end{array}
\end{equation}
From the results presented in Appendix~\ref{app:nlVSlor}, we can see
that the above operators give rise to $W^+W^-ZZ$, $ZZZZ$,
$\gamma Z Z Z$, $\gamma Z W^+ W^-$, and $\gamma \gamma Z Z$ anomalous
QGC.

\subsubsection{${\mathbf (D=3)^2}$ terms} 

We find 11 operators in this class with all of them respecting the
custodial symmetry.
\begin{equation}
\begin{array}{lclc}
{\cal T}^{p=6}_{34} = 
\Tr[V_\alpha {\cal D}_{\mu\nu}]\Tr[ V^\alpha{\cal D}^{\mu\nu}] 
&,
&{\cal T}^{p=6}_{35} = 
\Tr[ V_\alpha {\cal D}_{\mu\nu}]\Tr[ V^\nu{\cal D}^{\mu\alpha}] 
&,
\\
{\cal T}^{p=6}_{36} = 
\Tr[ V^\nu{\cal D}_{\mu\nu}]\Tr[ V_\alpha{\cal D}^{\mu\alpha}] 
&,
&{\cal T}^{p=6}_{37} = 
\Tr[ V_\alpha {\cal D}_{\mu}^\mu]\Tr[ V^\alpha{\cal D}^{\nu}_\nu] 
&,
\\
{\cal T}^{p=6}_{38} = 
\Tr[ V_{\nu} {\cal D}_{\mu}^\mu]\Tr[ V_\alpha{\cal D}^{\nu\alpha}] 
&,
&{\cal T}^{p=6}_{39} = i\, g \,
\Tr[ V_\alpha {\cal D}_{\mu\nu}]\Tr[ V^\nu \widehat{W}^{\mu\alpha}] 
&,
\\
{\cal T}^{p=6}_{40} = i\, g \,
\Tr[ V^\nu{\cal D}_{\mu\nu}]\Tr[ V_\alpha \widehat{W}^{\mu\alpha}] 
&,
&{\cal T}^{p=6}_{41} = i\, g \,
\Tr[ V_{\nu} {\cal D}_{\mu}^\mu]\Tr[ V_\alpha \widehat{W}^{\nu\alpha}] 
&,
\\
{\cal T}^{p=6}_{42} = -g^2\,
\Tr[V_\alpha \widehat{W}_{\mu\nu}]\Tr[ V^\alpha \widehat{W}^{\mu\nu}] 
&,
&{\cal T}^{p=6}_{43} = -g^2\,
\Tr[ V_\alpha \widehat{W}_{\mu\nu}]\Tr[ V^\nu \widehat{W}^{\mu\alpha}] 
&,
\\
{\cal T}^{p=6}_{44} = -g^2\,
\Tr[ V^\nu \widehat{W}_{\mu\nu}]\Tr[ V_\alpha \widehat{W}^{\mu\alpha}] 
&.
&&
\end{array}
\end{equation}
From Appendix~\ref{app:nlVSlor} we can learn that the effective
Lagrangians in this class generate $W^+W^-W^+W^-$, $W^+W^- ZZ$ and
$Z Z Z Z$ quartic vertices, as well as, $\gamma Z W^+ W^-$,
$\gamma Z Z Z $ and $\gamma \gamma Z Z$.  Due to the custodial
symmetry the last three vertices are multiplied by $s_w$, therefore,
vanishing in the custodial conserving limit $s_w \to 0$.

\subsubsection{${\mathbf (D=2)^3}$ terms}

There are 12 operators in this class
\begin{equation}
\begin{array}{lclc}
{\cal T}^{p=6}_{45} = \Tr[T {\cal D}_{\mu\nu}]\Tr[T {\cal D}^{\mu\nu}]
\Tr[V^\alpha V_\alpha] 
&,
&{\cal T}^{p=6}_{46} = \Tr[T {\cal D}_{\mu\nu}]\Tr[T {\cal D}^{\mu\alpha}]
\Tr[ V^\nu V_\alpha] 
&,
\\
{\cal T}^{p=6}_{47} = \Tr[T {\cal D}^\mu_{\mu}]\Tr[T {\cal D}^{\nu}_\nu]
\Tr[ V^\alpha V_\alpha] 
&,
&{\cal T}^{p=6}_{48} = \Tr[T {\cal D}^\mu_{\mu}]\Tr[T {\cal D}^{\nu\alpha}]
\Tr[ V_\nu V_\alpha] 
&,
\\
{\cal T}^{p=6}_{49} = -g^2\,\Tr[T \widehat{W}_{\mu\nu}]\Tr[T \widehat{W}^{\mu\nu}]
\Tr[ V^\alpha V_\alpha] 
&,
&{\cal T}^{p=6}_{50} = -g^2\,\Tr[T \widehat{W}_{\mu\nu}]\Tr[T \widehat{W}^{\mu\alpha}]
\Tr[ V^\nu  V_\alpha] 
&,
\\
{\cal T}^{p=6}_{51} = i\, g \,\Tr[T \widehat {W}_{\mu\nu}]\Tr[T {\cal D}^{\mu\alpha}]
\Tr[ V^\nu  V_\alpha] 
&,
&{\cal T}^{p=6}_{52} = -{g'}^2 \,{B}_{\mu\nu} {B}^{\mu\nu}
\Tr[ V^\alpha  V_\alpha] 
&,
\\
{\cal T}^{p=6}_{53} = -{g'}^2 \, B_{\mu\nu} B^{\mu\alpha}
\Tr[ V^\nu  V_\alpha] 
&,
&{\cal T}^{p=6}_{54} = i\, g' \,B_{\mu\nu}\Tr[T {\cal D}^{\mu\alpha}]
\Tr[ V^\nu  V_\alpha] 
&,
\\
{\cal T}^{p=6}_{55} = -g g'\,B_{\mu\nu}\Tr[T \widehat{W}^{\mu\nu}]
\Tr[ V^\alpha  V_\alpha] 
&,
&{\cal T}^{p=6}_{56} = -g g' \,B_{\mu\nu}\Tr[T \widehat{W}^{\mu\alpha}]
\Tr[ V^\nu  V_\alpha] 
&.
\end{array}
\end{equation}
It is interesting to notice that the operators in this group generate
QGC among all electroweak gauge bosons except for $W^+W^-W^+W^-$.

\subsubsection{${\mathbf (D=2)(D=4)}$ terms}

This class contains 7 operators
\begin{equation}
\begin{array}{lclc}
{\cal T}^{p=6}_{57} = \Tr[{\cal D}_{\mu\nu} {\cal D}^{\mu\nu}]
\Tr[V^\alpha V_\alpha] 
&,
&{\cal T}^{p=6}_{58} = \Tr[{\cal D}_{\mu\nu} {\cal D}^{\mu\alpha}]
\Tr[ V^\nu V_\alpha] 
&,
\\
{\cal T}^{p=6}_{59} = \Tr[{\cal D}^\mu_{\mu}  {\cal D}^{\nu}_\nu]
\Tr[ V^\alpha V_\alpha] 
&,
&{\cal T}^{p=6}_{60} = \Tr[{\cal D}^\mu_{\mu} {\cal D}^{\nu\alpha}]
\Tr[ V_\nu V_\alpha] 
&,
\\
{\cal T}^{p=6}_{61} = -g^2\, \Tr[ \widehat{W}_{\mu\nu}  \widehat{W}^{\mu\nu}]
\Tr[ V^\alpha V_\alpha] 
&,
&{\cal T}^{p=6}_{62} = -g^2\, \Tr[ \widehat{W}_{\mu\nu}  \widehat{W}^{\mu\alpha}]
\Tr[ V^\nu  V_\alpha] 
&,
\\
{\cal T}^{p=6}_{63} = i\, g\,\Tr[ \widehat {W}_{\mu\nu}  {\cal D}^{\mu\alpha}]
\Tr[ V^\nu  V_\alpha] 
&.
&&
\end{array}
\end{equation}
These operators are $SU(2)_c$ invariant in the limit $s_w \to 0$ and
this can be seen by their expression in terms of Lorentz structures
presented in Appendix~\ref{app:nlVSlor}. This class of Lagrangians
generate the following QGC: $W^+W^- ZZ$, $Z Z Z Z$,
$\gamma Z W^+ W^-$, $\gamma \gamma W^+ W^-$, $\gamma Z Z Z $, and
$\gamma \gamma Z Z$.

\subsubsection{${\mathbf (D=1)^2(D=4)}$ terms}

We find 7 operators in this class that violate the custodial symmetry:
\begin{equation}
\begin{array}{lclc}
{\cal T}^{p=6}_{64} = \Tr[{\cal D}_{\mu\nu} {\cal D}^{\mu\nu}]
\Tr[T V^\alpha] \Tr[T V_\alpha] 
&,
&{\cal T}^{p=6}_{65} = \Tr[{\cal D}_{\mu\nu} {\cal D}^{\mu\alpha}]
\Tr[T V^\nu] \Tr[TV_\alpha] 
&,
\\
{\cal T}^{p=6}_{66} = \Tr[{\cal D}^\mu_{\mu}  {\cal D}^{\nu}_\nu]
\Tr[T V^\alpha] \Tr[T V_\alpha] 
&,
&{\cal T}^{p=6}_{67} = \Tr[{\cal D}^\mu_{\mu} {\cal D}^{\nu\alpha}]
\Tr[T V_\nu] \Tr[T V_\alpha] 
&,
\\
{\cal T}^{p=6}_{68} = -g^2\, \Tr[ \widehat{W}_{\mu\nu}  \widehat{W}^{\mu\nu}]
\Tr[T V^\alpha] \Tr[T V_\alpha] 
&,
&{\cal T}^{p=6}_{69} = -g^2\, \Tr[ \widehat{W}_{\mu\nu}  \widehat{W}^{\mu\alpha}]
\Tr[T V^\nu ] \Tr[T V_\alpha] 
&,
\\
{\cal T}^{p=6}_{70} = i\, g\,\Tr[ \widehat {W}_{\mu\nu}  {\cal D}^{\mu\alpha}]
\Tr[T V^\nu] \Tr[T  V_\alpha] 
&.
&&
\end{array}
\end{equation}
As we can see in Appendix~\ref{app:nlVSlor}, this group of effective
Lagrangians give rise to $W^+W^- ZZ$, $Z Z Z Z$, $\gamma Z Z Z $, and
$\gamma \gamma Z Z$ QGC.

Altogether we find 70 independent operators leading to genuine QGC in
the chiral Lagrangian at ${\cal O}(p^6)$, so there are as many
operators as independent Lorentz structures containing two
derivatives.  As mentioned above, this was also the case at
${\cal O}(p^4)$. This is somehow not unexpected: as is well
known~\cite{Burgess:1992gx}, a generic $U(1)_{\rm em}$ invariant
Lagrangian, which is the only symmetry imposed in building the Lorentz
structures, is also invariant under nonlinear
$SU(2)_L \otimes U(1)_{\rm Y}$ transformations.

\section{Summary}
\label{sec:bounds}

In this work we have constructed the most general form of the QGC
containing up to two derivatives acting on the electroweak gauge boson
fields.  We have shown that there are 5 independent Lorentz structures
that respect the $U(1)_{\text em}$ symmetry and contain no derivatives
while there are 70 structures exhibiting two derivatives.

We have then derived which of these QGC are generated assuming that
the $SU(2)_L \otimes U(1)_Y$ gauge symmetry is linearly realized, as
characteristic of scenarios with a fundamental Higgs doublet. In this
case the lowest dimension that presents QGC without a TGC associated
to them is eight.  In this scenarios there are only three operators
that contain only massive gauge bosons and no derivative acting on
them; see Eq.~(\ref{eq:dphi}). So due to the linear realization of the
symmetry the Wilson coefficient of the five Lorentz structures that
contain no derivatives are correlated --independently of the basis of
operators used; see as example last line in Eq.~(\ref{eq:lin1}). In
the same framework, we find seven operators containing genuine QGC with
two derivatives and they generate only 23 of the 70 possible Lorentz
structures. So again, gauge invariance in the linear realization
implies correlations among the coefficients of the different Lorentz
structures, as for example those in Eq.~(\ref{eq:lin3}).

We also classified the quartic gauge-boson interactions assuming that
the $SU(2)_L \otimes U(1)_Y$ symmetry is realized nonlinearly with the
global symmetry breaking $SU(2)_L \otimes SU(2)_R \to SU(2)_c$,
characteristic of scenarios with a light dynamical Higgs boson.  At
order ${\cal O}(p^4)$ there are five chiral operators which generate
QGC without an associated TQC. They contain only $W^\pm$ and $Z$ and
no derivatives. There are 70 independent operators at order
${\cal O}(p^6)$ and they contain four gauge bosons and two
derivatives. This is, the chiral Lagrangian for genuine QGC contains
the same number of operators as independent Lorentz structures.  So no
basis independent correlation can be derived between the coefficients
of the Lorentz structures in this case.

At present the most sensitive searches for quartic gauge boson
couplings are those involving vertices with two photons.  Most of the
analyses carried out by the LEP~\cite{LEP:QGC, Achard:2002iz},
D0~\cite{Teva:QGC} and LHC~\cite{LHC:QGCvaa, LHC:QGCwaaEXC}
collaborations used the following effective Lagrangian to study the
two-photon sector~\cite{Belanger:1992qh}
\begin{equation}
{\cal L}_{\text eff} = -\frac{\alpha_{ \text em} \pi}{2}
\frac{a_0^W}{\Lambda^2} {\cal Q}^{\partial=2}_{\gamma W,1}
-\frac{\alpha_{\text em} \pi}{2}
\frac{a_c^W}{\Lambda^2} {\cal Q}^{\partial=2}_{\gamma W,2}
-\frac{\alpha_{ \text em} \pi}{2 c_w^2}
\frac{a_0^Z}{\Lambda^2} {\cal Q}^{\partial=2}_{\gamma Z,1}
-\frac{\alpha_{\text em} \pi}{2 c_w^2}
\frac{a_c^Z}{\Lambda^2} {\cal Q}^{\partial=2}_{\gamma Z,2}   \;,
\label{eq:lep_qgc}
\end{equation}
where $\alpha_{\text em}$ stands for the electromagnetic
fine-structure constant.  In the framework of electroweak gauge
invariance linearly realized, the seven operators in
Eq.~(\ref{eq:lind2}) give rise to QGC containing two photons which in
the notation in Eq.~(\ref{eq:lep_qgc}) read:
\begin{equation}
\begin{array}{ll}
   a_0^W  = - \frac{M_W^2}{\pi \alpha_{\text em}} \left[  
      s^2_w   \frac{f_{M,0}}{\Lambda^2}  +
  2 c_w^2  \frac{f_{M,2}}{\Lambda^2}  +
  s_w c_w \frac{f_{M,4}}{\Lambda^2}  
    \right]
&,
\\
   a_c^W = - \frac{M_W^2}{\pi \alpha_{\text em}} \left[  
- s_w^2  \frac{f_{M,1}}{\Lambda^2}
- c_w^2  \frac{f_{M,3}}{\Lambda^2}
+ 2 s_w c_w  \frac{f_{M,5}}{\Lambda^2}
+ \frac{s_w^2}{2}  \frac{f_{M,7}}{\Lambda^2}
\right]
&,
\\
   a_0^Z  = - \frac{M_W^2 c_w^2}{\pi \alpha_{\text em}} \left[  
      \frac{s^2_w}{2 c^2_w}   \frac{f_{M,0}}{\Lambda^2}  +
  \frac{f_{M,2}}{\Lambda^2}  
  -\frac{ s_w}{2 c_w} \frac{f_{M,4}}{\Lambda^2}  
    \right]
&,
\\
   a_c^Z = - \frac{M_W^2 c_w^2}{\pi \alpha_{\text em}} \left[  
- \frac{s_w^2}{2 c^2_w}  \frac{f_{M,1}}{\Lambda^2}
- \frac{1}{2}  \frac{f_{M,3}}{\Lambda^2}
- \frac{s_w }{c_w}  \frac{f_{M,5}}{\Lambda^2}
+ \frac{s_w^2}{4 c_w^2}  \frac{f_{M,7}}{\Lambda^2}
\right]
&.
\end{array}
\end{equation}
So even in the scenario with the linear realization of the gauge
invariance, these four coefficients can be fully uncorrelated.  Test
of the presence/absence of the correlations which can point out
towards an underlaying linear or chiral expansion will require the
measurement with equivalent accuracy of quartic vertices involving one
or zero photons, consequently requiring much more data.

\section*{Acknowledgments}

We would like thank Jean-Francois Laporte for asking questions that
lead to this work.  O.J.P.E. is supported in part by Conselho Nacional
de Desenvolvimento Cient\'{\i}fico e Tecnol\'ogico (CNPq) and by
Funda\c{c}\~ao de Amparo \`a Pesquisa do Estado de S\~ao Paulo
(FAPESP); M.C.G-G is supported by USA-NSF grant PHY-13-16617 and by
FP7 ITN INVISIBLES (Marie Curie Actions PITN-GA-2011-289442).  M.C.G-G
also acknowledges support by grants 2014-SGR-104 and by FPA2013-46570
and consolider-ingenio 2010 program CSD-2008-0037.

\appendix

\section{Relations between linear operators and Lorentz structures}
\label{app:linVSlor}

In the framework where the SM gauge symmetry is realized linearly, the
genuine quartic gauge couplings generated by the dimension-eight
operators with two derivatives listed in Section~\ref{sec:linear} can
be expressed in terms of the Lorentz structures defined in
Section~\ref{sec:lorentz} as:
\begin{eqnarray}
 {\cal O}_{M,0} 
 &=& \frac{g^2 v^2}{8} \left[ 
2{\cal{Q}}^{\partial=2}_{WW,1}
+ \frac{1}{c_w^2}{\cal{Q}}^{\partial=2}_{WZ,1}+ c_w^2{\cal{Q}}^{\partial=2}_{WZ,6}
+2 s_w c_w {\cal{Q}}^{\partial=2}_{\gamma ZW,1}
\right . \nonumber 
\\
&&\left. 
+ s_w^2{\cal{Q}}^{\partial=2}_{\gamma W,1}
+\frac{1}{2} {\cal{Q}}^{\partial=2}_{ZZ,1}+\frac{s_w}{c_w}{\cal{Q}}^{\partial=2}_{\gamma ZZ,1}
+\frac{s_w^2}{2 c_w^2}{\cal{Q}}^{\partial=2}_{\gamma Z,1}\right]
\\
 {\cal O}_{M,1} 
 &=& \frac{-g^2 v^2}{8} \left[ 
{\cal{Q}}^{\partial=2}_{WW,2}+{\cal{Q}}^{\partial=2}_{WW,3}
+ \frac{1}{c_w^2}{\cal{Q}}^{\partial=2}_{WZ,2}+ c_w^2{\cal{Q}}^{\partial=2}_{WZ,7}
+s_w c_w {\cal{Q}}^{\partial=2}_{\gamma ZW,2}
\right . \nonumber 
\\
&& \left. 
+ s_w^2{\cal{Q}}^{\partial=2}_{\gamma W,2}
+\frac{1}{2} {\cal{Q}}^{\partial=2}_{ZZ,2}+\frac{s_w}{c_w}{\cal{Q}}^{\partial=2}_{\gamma ZZ,2}
+\frac{s_w^2}{2 c_w^2}{\cal{Q}}^{\partial=2}_{\gamma Z,2}\right]
\\
 {\cal O}_{M,2} 
 &=& \frac{g^2 v^2}{4} \left[
 c_w^2{\cal{Q}}^{\partial=2}_{\gamma W,1}-2 s_w c_w {\cal{Q}}^{\partial=2}_{\gamma ZW,1}
+s_w^2{\cal{Q}}^{\partial=2}_{WZ,6} 
\right . \nonumber 
\\
&& \left. 
+\frac{1}{2} {\cal{Q}}^{\partial=2}_{\gamma Z,1}-\frac{s_w}{c_w}{\cal{Q}}^{\partial=2}_{\gamma ZZ,1}
+\frac{s_w^2}{2 c_w^2}{\cal{Q}}^{\partial=2}_{ZZ,1}\right]
\\
 {\cal O}_{M,3} 
 &=& \frac{-g^2 v^2}{8} \left[
 c_w^2{\cal{Q}}^{\partial=2}_{\gamma W,2}- s_w c_w {\cal{Q}}^{\partial=2}_{\gamma ZW ,2}
+s_w^2{\cal{Q}}^{\partial=2}_{WZ,7} 
\right . \nonumber \\
&& \left. 
+\frac{1}{2} {\cal{Q}}^{\partial=2}_{\gamma Z,2}-\frac{s_w}{c_w}{\cal{Q}}^{\partial=2}_{\gamma ZZ,2}
+\frac{s_w^2}{2 c_w^2}{\cal{Q}}^{\partial=2}_{ZZ,2}\right]
\\
 {\cal O}_{M,4} 
 &=& \frac{g^2 v^2}{8} \left[(c_w^2-s_w^2)
\left({\cal{Q}}^{\partial=2}_{\gamma ZW,1}
-\frac{1}{2c_w^2} {\cal{Q}}^{\partial=2}_{\gamma ZZ,1}\right)
-{\cal{Q}}^{\partial=2}_{\gamma ZW,3}+\frac{s_w}{c_w}{\cal{Q}}^{\partial=2}_{WZ,3}
\right . \nonumber \\
&& \left. 
+s_w c_w
\left({\cal{Q}}^{\partial=2}_{\gamma W,1}-{\cal{Q}}^{\partial=2}_{WZ,6} -
\frac{1}{2 c_w^2} ({\cal{Q}}^{\partial=2}_{\gamma Z,1}-{\cal{Q}}^{\partial=2}_{ZZ,1}
\right)
\right]
\\
 {\cal O}_{M,5} 
 &=& \frac{g^2 v^2}{8} \left[(c_w^2-s_w^2)
\left({\cal{Q}}^{\partial=2}_{\gamma ZW,2}
-\frac{1}{c_w^2} {\cal{Q}}^{\partial=2}_{\gamma ZZ,2}\right)
\right . \nonumber \\
&& \left. 
-{\cal{Q}}^{\partial=2}_{\gamma ZW,4}-{\cal{Q}}^{\partial=2}_{\gamma ZW,5}
+\frac{s_w}{c_w}({\cal{Q}}^{\partial=2}_{WZ,4}+{\cal{Q}}^{\partial=2}_{WZ,5})
\right . \nonumber \\
&& \left. 
+s_w c_w
\left(2({\cal{Q}}^{\partial=2}_{\gamma W,2}-{\cal{Q}}^{\partial=2}_{WZ,7})-
\frac{1}{ c_w^2} ({\cal{Q}}^{\partial=2}_{\gamma Z,2}
-{\cal{Q}}^{\partial=2}_{ZZ,2}
\right)
\right]
\\
 {\cal O}_{M,7} 
 &=& \frac{g^2 v^2}{16} \left[
2 {\cal{Q}}^{\partial=2}_{WW,3}
+ c_w^2 {\cal{Q}}^{\partial=2}_{WZ,7}
+ s_w c_w  {\cal{Q}}^{\partial=2}_{\gamma Z W,2}
+s_w^2 {\cal{Q}}^{\partial=2}_{\gamma W,2}
\right . \nonumber \\
&& \left. -({\cal{Q}}^{\partial=2}_{WZ,5}-{\cal{Q}}^{\partial=2}_{WZ,4})
-\frac{s_w}{c_w}({\cal{Q}}^{\partial=2}_{\gamma Z W,5}
-{\cal{Q}}^{\partial=2}_{\gamma Z W,4})
\right . \nonumber \\
&& \left. 
+\frac{1}{2} {\cal{Q}}^{\partial=2}_{ZZ,2}
+\frac{s_w}{c_w} {\cal{Q}}^{\partial=2}_{\gamma ZZ,2}
+\frac{s^2_w}{2c^2_w} {\cal{Q}}^{\partial=2}_{\gamma Z,2} 
+\frac{1}{c_w^2} {\cal{Q}}^{\partial=2}_{WZ,2} 
\right]
\end{eqnarray}

\section{Dimension-eight operators containing four field strength tensors}
\label{ap:4FS}

There are 8 operators containing just field strength tensors that lead
to genuine quartic anomalous couplings, which in the notation used in
Ref.~\cite{Eboli:2006wa} are:
\begin{equation}
\begin{array} {lcl}
 {\cal O}_{T,0} =   \hbox{Tr}\left [ \widehat{W}_{\mu\nu} \widehat{W}^{\mu\nu} \right ]
\times   \hbox{Tr}\left [ \widehat{W}_{\alpha\beta} \widehat{W}^{\alpha\beta} \right ]
&, &
 {\cal O}_{T,1} =   \hbox{Tr}\left [ \widehat{W}_{\alpha\nu} \widehat{W}^{\mu\beta} \right ]
\times   \hbox{Tr}\left [ \widehat{W}_{\mu\beta} \widehat{W}^{\alpha\nu} \right ]
\\
 {\cal O}_{T,2} =   \hbox{Tr}\left [ \widehat{W}_{\alpha\mu} \widehat{W}^{\mu\beta} \right ]
\times   \hbox{Tr}\left [ \widehat{W}_{\beta\nu} \widehat{W}^{\nu\alpha} \right ]
&,&
 {\cal O}_{T,5} =   \hbox{Tr}\left [ \widehat{W}_{\mu\nu} \widehat{W}^{\mu\nu} \right ]
\times   B_{\alpha\beta} B^{\alpha\beta}
\\ 
 {\cal O}_{T,6} =   \hbox{Tr}\left [ \widehat{W}_{\alpha\nu} \widehat{W}^{\mu\beta} \right ]
\times   B_{\mu\beta} B^{\alpha\nu} 
&,&
 {\cal O}_{T,7} =   \hbox{Tr}\left [ \widehat{W}_{\alpha\mu} \widehat{W}^{\mu\beta} \right ]
\times   B_{\beta\nu} B^{\nu\alpha} 
\\ 
 {\cal O}_{T,8} =   B_{\mu\nu} B^{\mu\nu}  B_{\alpha\beta} B^{\alpha\beta}
&,& 
 {\cal O}_{T,9} =  B_{\alpha\mu} B^{\mu\beta}   B_{\beta\nu} B^{\nu\alpha} 
\; . 
\end{array}
\end{equation}
%


\section{Relations between ${\cal O}(p^6)$ 
Operators and Lorentz Structures}
\label{app:nlVSlor}

Expanding the 70 ${\cal O}(p^6)$  operators introduced in Sec.~\ref{sec:chiral}
in terms of the Lorentz structures in Sec.~\ref{sec:lorentz2d} we find
\begin{eqnarray}
{\cal T}^{p=6}_1 &=& \frac{g^4}{c_w^4}{\cal{Q}}^{\partial=2}_{ZZ,3} \\
{\cal T}^{p=6}_2 &=& \frac{g^4}{c_w^4}{\cal{Q}}^{\partial=2}_{ZZ,4} \\
{\cal T}^{p=6}_3 &=& \frac{g^4}{c_w^4}{\cal{Q}}^{\partial=2}_{ZZ,5} \\
{\cal T}^{p=6}_4 &=& \frac{g^4}{c_w^4}{\cal{Q}}^{\partial=2}_{ZZ,6} \\
{\cal T}^{p=6}_5 &=& \frac{g^4}{c_w^2}
\left[c_w^2 {\cal{Q}}^{\partial=2}_{ZZ,1} +2 s_w c_w  {\cal{Q}}^{\partial=2}_{\gamma ZZ,1}+s_w^2 {\cal{Q}}^{\partial=2}_{\gamma Z,1}\right]
\\
{\cal T}^{p=6}_6 &=& \frac{g^4}{c_w^2}
\left[c_w^2 {\cal{Q}}^{\partial=2}_{ZZ,2} +2 s_w c_w  {\cal{Q}}^{\partial=2}_{\gamma ZZ,2}+s_w^2 {\cal{Q}}^{\partial=2}_{\gamma Z,2}\right]
\\
{\cal T}^{p=6}_7 &=& \frac{g^4}{c_w^3}
\left[c_w {\cal{Q}}^{\partial=2}_{ZZ,7} + s_w   {\cal{Q}}^{\partial=2}_{\gamma ZZ,3}\right]
\\
{\cal T}^{p=6}_8 &=& \frac{g^2{g'}^2 }{c_w^2}
\left[s_w^2 {\cal{Q}}^{\partial=2}_{ZZ,1} -2 s_w c_w  {\cal{Q}}^{\partial=2}_{\gamma ZZ,1}+c_w^2 {\cal{Q}}^{\partial=2}_{\gamma Z,1}\right]
\\
{\cal T}^{p=6}_9 &=& \frac{g^2{g'}^2 }{c_w^2}
\left[s_w^2 {\cal{Q}}^{\partial=2}_{ZZ,2} -2 s_w c_w  {\cal{Q}}^{\partial=2}_{\gamma ZZ,2}+c_w^2 {\cal{Q}}^{\partial=2}_{\gamma Z,2}\right]
\\
{\cal T}^{p=6}_{10} &=& \frac{g^3{g'} }{c_w^3}
\left[-s_w {\cal{Q}}^{\partial=2}_{ZZ,7} +c_w  {\cal{Q}}^{\partial=2}_{\gamma ZZ,3}\right]
\\
{\cal T}^{p=6}_{11} &=& \frac{g^3{g'} }{c_w^2}
\left[c_w s_w \left(-{\cal{Q}}^{\partial=2}_{ZZ,1} +{\cal{Q}}^{\partial=2}_{\gamma Z,1} \right)
+(c_w^2-s_w^2)  {\cal{Q}}^{\partial=2}_{\gamma ZZ,1}\right]
 \\
{\cal T}^{p=6}_{12} &=& \frac{g^3{g'} }{c_w^2}
\left[c_w s_w \left(-{\cal{Q}}^{\partial=2}_{ZZ,2} +{\cal{Q}}^{\partial=2}_{\gamma Z,2} \right)
+(c_w^2-s_w^2)  {\cal{Q}}^{\partial=2}_{\gamma ZZ,2}\right]
\\
%
{\cal T}^{p=6}_{13} &=& \frac{g^4}{2 c_w^2}\left[
{\cal{Q}}^{\partial=2}_{W Z,12}+ {\frac{1}{c_w^2} }
{\cal{Q}}^{\partial=2}_{ZZ,3}\right]
\\
{\cal T}^{p=6}_{14} &=& \frac{g^4}{2 c_w^2}\left[
{\cal{Q}}^{\partial=2}_{WZ,14}+ {\frac{1}{c_w^2} }
{\cal{Q}}^{\partial=2}_{ZZ,4}\right]
\\
{\cal T}^{p=6}_{15} &=& \frac{g^4}{2 c_w^2}\left[
{\cal{Q}}^{\partial=2}_{WZ,13}+ {\frac{1}{c_w^2} }
{\cal{Q}}^{\partial=2}_{ZZ,4}\right]
\\
{\cal T}^{p=6}_{16} &=& \frac{g^4}{2 c_w^2}\left[
{\cal{Q}}^{\partial=2}_{WZ,15}+ {\frac{1}{c_w^2} }
{\cal{Q}}^{\partial=2}_{ZZ,5}\right]
\\
{\cal T}^{p=6}_{17} &=& \frac{g^4}{2 c_w^2}\left[
{\cal{Q}}^{\partial=2}_{WZ,17}+{\frac{1}{c_w^2} }
{\cal{Q}}^{\partial=2}_{ZZ,6}\right]
\\
{\cal T}^{p=6}_{18} &=& \frac{g^4}{2 c_w^2}\left[
{\cal{Q}}^{\partial=2}_{WZ,16}+ {\frac{1}{c_w^2} }
{\cal{Q}}^{\partial=2}_{ZZ,6}\right]
\\
{\cal T}^{p=6}_{19} &=& \frac{g^4}{2 c_w^2}\left[
{\cal{Q}}^{\partial=2}_{WZ,23}+{\cal{Q}}^{\partial=2}_{ZZ,7}+ \frac{s_w}{c_w}{\cal{Q}}^{\partial=2}_{\gamma ZZ,3}
\right]
\\
{\cal T}^{p=6}_{20} &=& \frac{g^4}{2 c_w^2}\left[
{\cal{Q}}^{\partial=2}_{WZ,24}+{\cal{Q}}^{\partial=2}_{ZZ,7}+ \frac{s_w}{c_w}{\cal{Q}}^{\partial=2}_{\gamma ZZ,3}
\right]
\\
{\cal T}^{p=6}_{21} &=& \frac{g^4}{2 c_w^2} 
{\cal{Q}}^{\partial=2}_{WZ,25}  
\\
{\cal T}^{p=6}_{22} &=& \frac{g^4}{2 c_w}
\left[c_w\left({\cal{Q}}^{\partial=2}_{WZ,26}+\frac{1}{c_w^2}{\cal{Q}}^{\partial=2}_{ZZ,7}\right)+ 
s_w\left({\cal{Q}}^{\partial=2}_{\gamma ZW,7} +\frac{1}{c_w^2}{\cal{Q}}^{\partial=2}_{\gamma ZZ,3}  \right)\right]
\\
{\cal T}^{p=6}_{23} &=& \frac{g^4}{2 c_w}
\left[c_w\left({\cal{Q}}^{\partial=2}_{WZ,27}+\frac{1}{c_w^2}{\cal{Q}}^{\partial=2}_{ZZ,7}\right)+ 
s_w\left({\cal{Q}}^{\partial=2}_{\gamma ZW,8}+\frac{1}{c_w^2}{\cal{Q}}^{\partial=2}_{\gamma ZZ,3}  \right)\right]
\\
{\cal T}^{p=6}_{24} &=& \frac{g^4}{2 c_w}
\left[c_w {\cal{Q}}^{\partial=2}_{WZ,28}+s_w {\cal{Q}}^{\partial=2}_{\gamma ZW,9} \right]
\\
{\cal T}^{p=6}_{25} &=& \frac{g^4}{2 c_w}
\left[
c_w\left({\cal{Q}}^{\partial=2}_{WZ,3}+{\cal{Q}}^{\partial=2}_{ZZ,1}+\frac{s_w}{c_w}
{\cal{Q}}^{\partial=2}_{\gamma ZZ,1}\right) 
+s_w\left({\cal{Q}}^{\partial=2}_{\gamma ZW,3}+{\cal{Q}}^{\partial=2}_{\gamma ZZ,1}+\frac{s_w}{c_w}
{\cal{Q}}^{\partial=2}_{\gamma Z,1}\right)\right]
\\
{\cal T}^{p=6}_{26} &=& \frac{g^4}{2 c_w}
\left[
c_w\left({\cal{Q}}^{\partial=2}_{WZ,4}+{\cal{Q}}^{\partial=2}_{ZZ,2}+\frac{s_w}{c_w}
{\cal{Q}}^{\partial=2}_{\gamma ZZ,2}\right)
+s_w\left({\cal{Q}}^{\partial=2}_{\gamma ZW,4}+{\cal{Q}}^{\partial=2}_{\gamma ZZ,2}+\frac{s_w}{c_w}
{\cal{Q}}^{\partial=2}_{\gamma Z,2}\right)\right]
\\
{\cal T}^{p=6}_{27} &=& \frac{g^4}{2 c_w}
\left[
c_w\left({\cal{Q}}^{\partial=2}_{WZ,5}+{\cal{Q}}^{\partial=2}_{ZZ,2}+\frac{s_w}{c_w}
{\cal{Q}}^{\partial=2}_{\gamma ZZ,2}\right)
+s_w\left({\cal{Q}}^{\partial=2}_{\gamma ZW,5}+{\cal{Q}}^{\partial=2}_{\gamma ZZ,2}+\frac{s_w}{c_w}
{\cal{Q}}^{\partial=2}_{\gamma Z,2}\right)\right]
\\
{\cal T}^{p=6}_{28} &=& \frac{g^3 g'}{2 c_w}
\left[-s_w\left({\cal{Q}}^{\partial=2}_{WZ,26}+\frac{1}{c_w^2}{\cal{Q}}^{\partial=2}_{ZZ,7}\right)+ 
c_w\left({\cal{Q}}^{\partial=2}_{\gamma ZW,7} +\frac{1}{c_w^2}{\cal{Q}}^{\partial=2}_{\gamma ZZ,3}  \right)\right]
\\
{\cal T}^{p=6}_{29} &=& \frac{g^3 g'}{2 c_w}
\left[-s_w\left({\cal{Q}}^{\partial=2}_{WZ,27}+\frac{1}{c_w^2}{\cal{Q}}^{\partial=2}_{ZZ,7}\right)+ 
c_w\left({\cal{Q}}^{\partial=2}_{\gamma ZW,8}+\frac{1}{c_w^2}{\cal{Q}}^{\partial=2}_{\gamma ZZ,3}  \right)\right]
\\
{\cal T}^{p=6}_{30} &=& \frac{g^3 g'}{2 c_w}
\left[-s_w {\cal{Q}}^{\partial=2}_{WZ,28} + c_w {\cal{Q}}^{\partial=2}_{\gamma ZW,9} \right]
\\
{\cal T}^{p=6}_{31} &=& \frac{g^3g'}{2 c_w}
\left[
-s_w\left({\cal{Q}}^{\partial=2}_{WZ,3}+{\cal{Q}}^{\partial=2}_{ZZ,1}+\frac{s_w}{c_w}
{\cal{Q}}^{\partial=2}_{\gamma ZZ,1}\right)
+c_w\left({\cal{Q}}^{\partial=2}_{\gamma ZW,3}+{\cal{Q}}^{\partial=2}_{\gamma ZZ,1}+\frac{s_w}{c_w}
{\cal{Q}}^{\partial=2}_{\gamma Z,1}\right)\right]
\\
{\cal T}^{p=6}_{32} &=& \frac{g^3 g'}{2 c_w}
\left[
-s_w\left({\cal{Q}}^{\partial=2}_{WZ,4}+{\cal{Q}}^{\partial=2}_{ZZ,2}+\frac{s_w}{c_w}
{\cal{Q}}^{\partial=2}_{\gamma ZZ,2}\right)
+c_w\left({\cal{Q}}^{\partial=2}_{\gamma ZW,4}+{\cal{Q}}^{\partial=2}_{\gamma ZZ,2}+\frac{s_w}{c_w}
{\cal{Q}}^{\partial=2}_{\gamma Z,2}\right)\right]
\\
{\cal T}^{p=6}_{33} &=& \frac{g^3 g'}{2 c_w}
\left[
-s_w\left({\cal{Q}}^{\partial=2}_{WZ,5}+{\cal{Q}}^{\partial=2}_{ZZ,2}+\frac{s_w}{c_w}
{\cal{Q}}^{\partial=2}_{\gamma ZZ,2}\right)
+c_w\left({\cal{Q}}^{\partial=2}_{\gamma ZW,5}+{\cal{Q}}^{\partial=2}_{\gamma ZZ,2}+\frac{s_w}{c_w}
{\cal{Q}}^{\partial=2}_{\gamma Z,2}\right)\right]
\\
%
{\cal T}^{p=6}_{34} &=& \frac{g^4}{4 }\left[
{\cal{Q}}^{\partial=2}_{WW,11}+2{\cal{Q}}^{\partial=2}_{WW,6} +\frac{2}{c_w^2}{\cal{Q}}^{\partial=2}_{WZ,12}
+\frac{1}{c_w^4}{\cal{Q}}^{\partial=2}_{ZZ,3}\right] 
\\
{\cal T}^{p=6}_{35} &=& \frac{g^4}{4 }\left[
{\cal{Q}}^{\partial=2}_{WW,12}+ {\cal{Q}}^{\partial=2}_{WW,7}+{\cal{Q}}^{\partial=2}_{WW,8} 
+\frac{2}{c_w^2} {\cal{Q}}^{\partial=2}_{WZ,14}
+\frac{1}{c_w^4}{\cal{Q}}^{\partial=2}_{ZZ,4}\right] 
\\
{\cal T}^{p=6}_{36} &=& \frac{g^4}{4 }\left[
{\cal{Q}}^{\partial=2}_{WW,12}+{2} {\cal{Q}}^{\partial=2}_{WW,7}+  
\frac{2}{c_w^2}{\cal{Q}}^{\partial=2}_{WZ,13}
+\frac{1}{c_w^4}{\cal{Q}}^{\partial=2}_{ZZ,4}\right] 
\\
{\cal T}^{p=6}_{37} &=& \frac{g^4}{4 }\left[
{\cal{Q}}^{\partial=2}_{WW,13}+2{\cal{Q}}^{\partial=2}_{WW,9} +\frac{2}{c_w^2}{\cal{Q}}^{\partial=2}_{WZ,15}
+\frac{1}{c_w^4}{\cal{Q}}^{\partial=2}_{ZZ,5}\right] 
\\
{\cal T}^{p=6}_{38} &=& \frac{g^4}{4 }\left[
{\cal{Q}}^{\partial=2}_{WW,14}+{\cal{Q}}^{\partial=2}_{WW,10}+\frac{1}{c_w^2}{\cal{Q}}^{\partial=2}_{WZ,16} +
\frac{1}{c_w^2}
{\cal{Q}}^{\partial=2}_{WZ,17}
+\frac{1}{c_w^4}{\cal{Q}}^{\partial=2}_{ZZ,6}\right] 
\\
{\cal T}^{p=6}_{39} &=& \frac{g^4}{4 }\left[
{\cal{Q}}^{\partial=2}_{WW,18}+{\cal{Q}}^{\partial=2}_{WW,{16}}+{\cal{Q}}^{\partial=2}_{WZ,26} +
\frac{1}{c_w^2}{\cal{Q}}^{\partial=2}_{WZ,23}
+\frac{1}{c_w^2}{\cal{Q}}^{\partial=2}_{ZZ,7} 
+ \frac{s_w}{c_w}\left({\cal{Q}}^{\partial=2}_{\gamma ZW,7}+
\frac{1}{c_w^2}{\cal{Q}}^{\partial=2}_{\gamma ZZ,3}\right)
\right] 
\\
{\cal T}^{p=6}_{40} &=& \frac{g^4}{4 }\left[
{\cal{Q}}^{\partial=2}_{WW,18}+{\cal{Q}}^{\partial=2}_{WW,{15}}+{\cal{Q}}^{\partial=2}_{WZ,27} +
\frac{1}{c_w^2}{\cal{Q}}^{\partial=2}_{WZ,24}
+\frac{1}{c_w^2}{\cal{Q}}^{\partial=2}_{ZZ,7} 
+ \frac{s_w}{c_w}\left({\cal{Q}}^{\partial=2}_{\gamma ZW,8}+
\frac{1}{c_w^2}{\cal{Q}}^{\partial=2}_{\gamma ZZ,3}\right)
\right] 
\\
{\cal T}^{p=6}_{41} &=& \frac{g^4}{4 }\left[
- {\cal{Q}}^{\partial=2}_{WW,17} {-} {\cal{Q}}^{\partial=2}_{WZ,28} +
\frac{1}{c_w^2}{\cal{Q}}^{\partial=2}_{WZ,25}
{-}\frac{s_w}{c_w}{\cal{Q}}^{\partial=2}_{\gamma ZW,9}\right] 
 \\
{\cal T}^{p=6}_{42} &=& \frac{g^4}{4 }\left[
2{\cal{Q}}^{\partial=2}_{WW,1}+{\cal{Q}}^{\partial=2}_{WW,4}+2
                        {\cal{Q}}^{\partial=2}_{WZ,3}
                        +{\cal{Q}}^{\partial=2}_{ZZ,1} 
{+} 2\frac{s_w}{c_w}{\cal{Q}}^{\partial=2}_{\gamma ZZ,1}
+2\frac{s_w}{c_w}{\cal{Q}}^{\partial=2}_{\gamma ZW,3}
+ 
\frac{s_w^2}{c_w^2}{\cal{Q}}^{\partial=2}_{\gamma Z,1}
\right] 
 \\
{\cal T}^{p=6}_{43} &=& \frac{g^4}{4 }\left[
{\cal{Q}}^{\partial=2}_{WW,5}
+ {2} {\cal{Q}}^{\partial=2}_{WW,3}
+2 {\cal{Q}}^{\partial=2}_{WZ,4} +{\cal{Q}}^{\partial=2}_{ZZ,2} 
{+} 2\frac{s_w}{c_w}{\cal{Q}}^{\partial=2}_{\gamma ZZ,2}+ 
2\frac{s_w}{c_w}{\cal{Q}}^{\partial=2}_{\gamma ZW,4}
+\frac{s_w^2}{c_w^2}{\cal{Q}}^{\partial=2}_{\gamma Z,2}
\right] 
\\
{\cal T}^{p=6}_{44} &=& \frac{g^4}{4 }\left[
{\cal{Q}}^{\partial=2}_{WW,5}+{2} {\cal{Q}}^{\partial=2}_{WW,2} 
+2 {\cal{Q}}^{\partial=2}_{WZ,5} +{\cal{Q}}^{\partial=2}_{ZZ,2} 
{+} 2\frac{s_w}{c_w}{\cal{Q}}^{\partial=2}_{\gamma ZZ,2}
+2\frac{s_w}{c_w}{\cal{Q}}^{\partial=2}_{\gamma ZW,5}
+ \frac{s_w^2}{c_w^2}{\cal{Q}}^{\partial=2}_{\gamma Z,2}
\right] 
\\
%
{\cal T}^{p=6}_{45} &=& \frac{g^4}{2 c_w^2 }\left[
2{\cal{Q}}^{\partial=2}_{WZ,18}+\frac{1}{c_w^2}{\cal{Q}}^{\partial=2}_{ZZ,3}\right]
\\
{\cal T}^{p=6}_{46} &=& \frac{g^4}{2 c_w^2 }\left[
2{\cal{Q}}^{\partial=2}_{WZ,19}+\frac{1}{c_w^2}{\cal{Q}}^{\partial=2}_{ZZ,4}\right]
\\
{\cal T}^{p=6}_{47} &=& \frac{g^4}{2 c_w^2 }\left[
2{\cal{Q}}^{\partial=2}_{WZ,20}+\frac{1}{c_w^2}{\cal{Q}}^{\partial=2}_{ZZ,5}\right]
\\
{\cal T}^{p=6}_{48} &=& \frac{g^4}{2 c_w^2 }\left[
2{\cal{Q}}^{\partial=2}_{WZ,21}+\frac{1}{c_w^2}{\cal{Q}}^{\partial=2}_{ZZ,6}\right]
\\
{\cal T}^{p=6}_{49} &=& \frac{g^4}{2}\left[
2c_w^2{\cal{Q}}^{\partial=2}_{WZ,6}+4 s_w c_w {\cal{Q}}^{\partial=2}_{\gamma ZW,1}+
2 s_w^2 {\cal{Q}}^{\partial=2}_{\gamma W,1} 
+ {\cal{Q}}^{\partial=2}_{ZZ,1}+ 
2\frac{s_w}{c_w}{\cal{Q}}^{\partial=2}_{\gamma ZZ,1}
+\frac{s_w^2}{c_w^2}{\cal{Q}}^{\partial=2}_{\gamma Z,1} \right]
 \\
{\cal T}^{p=6}_{50} &=& \frac{g^4}{2}\left[
2c_w^2{\cal{Q}}^{\partial=2}_{WZ,7}+ {2} s_w c_w {\cal{Q}}^{\partial=2}_{\gamma ZW,2}+
2 s_w^2 {\cal{Q}}^{\partial=2}_{\gamma W,2} 
+ {\cal{Q}}^{\partial=2}_{ZZ,2}+ 
2\frac{s_w}{c_w}{\cal{Q}}^{\partial=2}_{\gamma ZZ,2}
+\frac{s_w^2}{c_w^2}{\cal{Q}}^{\partial=2}_{\gamma Z,2} \right]
\\
{\cal T}^{p=6}_{51} &=& \frac{g^4}{2 c_w}\left[
c_w{\cal{Q}}^{\partial=2}_{WZ,29}+ \frac{1}{c_w} {\cal{Q}}^{\partial=2}_{ZZ,7}+
s_w {\cal{Q}}^{\partial=2}_{\gamma Z W,6} +\frac{s_w}{c_w^2} {\cal{Q}}^{\partial=2}_{\gamma Z Z,3} 
\right]
\\
{\cal T}^{p=6}_{52} &=& \frac{g^2 {g'}^2}{2 }\left[
2 s_w^2{\cal{Q}}^{\partial=2}_{WZ,6}-4 s_w c_w {\cal{Q}}^{\partial=2}_{\gamma ZW,1}+
2 c_w^2 {\cal{Q}}^{\partial=2}_{\gamma W,1} 
+ \frac{s_w^2}{c_w^2}{\cal{Q}}^{\partial=2}_{ZZ,1} 
-2\frac{s_w}{c_w}{\cal{Q}}^{\partial=2}_{\gamma ZZ,1}
+{\cal{Q}}^{\partial=2}_{\gamma Z,1} \right]
 \\
{\cal T}^{p=6}_{53} &=& \frac{g^2 {g'}^2}{2 }\left[
2 s_w^2{\cal{Q}}^{\partial=2}_{WZ,7}- {2} s_w c_w {\cal{Q}}^{\partial=2}_{\gamma ZW,2}+
2 c_w^2 {\cal{Q}}^{\partial=2}_{\gamma W,2} 
+ \frac{s_w^2}{c_w^2}{\cal{Q}}^{\partial=2}_{ZZ,2} 
-2\frac{s_w}{c_w}{\cal{Q}}^{\partial=2}_{\gamma ZZ,2}
+{\cal{Q}}^{\partial=2}_{\gamma Z,2} \right]
 \\
{\cal T}^{p=6}_{54} &=& {\frac{g^3 {g'}}{2c_w}}\left[
-s_w{\cal{Q}}^{\partial=2}_{WZ,29}- \frac{s_w}{c_w^2} {\cal{Q}}^{\partial=2}_{ZZ,7}+
c_w {\cal{Q}}^{\partial=2}_{\gamma Z W,6} +\frac{1}{c_w} {\cal{Q}}^{\partial=2}_{\gamma Z Z,3} 
\right]
\\
{\cal T}^{p=6}_{55} &=& \frac{g^3 {g'}}{2}\left[
2(c_w^2-s_w^2) {\cal{Q}}^{\partial=2}_{\gamma Z W,1}+2 s_w c_w
\left({\cal{Q}}^{\partial=2}_{\gamma W,1}-{\cal{Q}}^{\partial=2}_{WZ,6}\right)
+ \frac{(c_w^2-s_w^2)}{c_w^2} {\cal{Q}}^{\partial=2}_{\gamma Z Z,1}+ \frac{s_w}{ c_w}
\left({\cal{Q}}^{\partial=2}_{\gamma Z,1}-{\cal{Q}}^{\partial=2}_{ZZ,1}\right) \right]
\\
{\cal T}^{p=6}_{56} &=& \frac{g^3 {g'}}{2}\left[
{(c_w^2-s_w^2)} {\cal{Q}}^{\partial=2}_{\gamma Z W,2}+2 s_w c_w
\left({\cal{Q}}^{\partial=2}_{\gamma W,2}-{\cal{Q}}^{\partial=2}_{WZ,7}\right)
+ \frac{(c_w^2-s_w^2)}{c_w^2} {\cal{Q}}^{\partial=2}_{\gamma Z Z,2}+ \frac{s_w}{ c_w}
\left({\cal{Q}}^{\partial=2}_{\gamma Z,2}-{\cal{Q}}^{\partial=2}_{ZZ,2}\right) \right]
\\
%
{\cal T}^{p=6}_{57} &=& \frac{g^4}{4 }\left[
4{\cal{Q}}^{\partial=2}_{WW,6}+\frac{2}{c_w^2}{\cal{Q}}^{\partial=2}_{WZ,8}
+\frac{2}{c_w^2} {\cal{Q}}^{\partial=2}_{WZ,18} +\frac{1}{c_w^4}
{\cal{Q}}^{\partial=2}_{ZZ,3} \right] 
\\
{\cal T}^{p=6}_{58} &=& \frac{g^4}{4 }\left[
2{\cal{Q}}^{\partial=2}_{WW,7}+2{\cal{Q}}^{\partial=2}_{WW,8}
+\frac{2}{c_w^2}{\cal{Q}}^{\partial=2}_{WZ,9}
+\frac{2}{c_w^2} {\cal{Q}}^{\partial=2}_{WZ,19} +\frac{1}{c_w^4}
{\cal{Q}}^{\partial=2}_{ZZ,4} \right] 
\\
{\cal T}^{p=6}_{59} &=& \frac{g^4}{4 }\left[
4{\cal{Q}}^{\partial=2}_{WW,9}+\frac{2}{c_w^2}{\cal{Q}}^{\partial=2}_{WZ,10}
+\frac{2}{c_w^2} {\cal{Q}}^{\partial=2}_{WZ,20} +\frac{1}{c_w^4}
{\cal{Q}}^{\partial=2}_{ZZ,5} \right] 
\\
{\cal T}^{p=6}_{60} &=& \frac{g^4}{4 }\left[
2{\cal{Q}}^{\partial=2}_{WW,10}+\frac{2}{c_w^2}{\cal{Q}}^{\partial=2}_{WZ,11}
+\frac{2}{c_w^2} {\cal{Q}}^{\partial=2}_{WZ,21} +\frac{1}{c_w^4}
{\cal{Q}}^{\partial=2}_{ZZ,6} \right] 
\\
{\cal T}^{p=6}_{61} &=& \frac{g^4}{4 }\left[
4{\cal{Q}}^{\partial=2}_{WW,1}+ 2 c_w^2{\cal{Q}}^{\partial=2}_{WZ,6}
+4 s_w c_w {\cal{Q}}^{\partial=2}_{\gamma ZW,1} + 2 s_w^2 {\cal{Q}}^{\partial=2}_{\gamma W,1}
\right. \nonumber \\ & & \left.
+ \frac{2}{c_w^2} {\cal{Q}}^{\partial=2}_{WZ,1}+ {\cal{Q}}^{\partial=2}_{ZZ,1}
+ \frac{2 s_w}{c_w} {\cal{Q}}^{\partial=2}_{\gamma ZZ,1}
+ \frac{s^2_w}{c^2_w} {\cal{Q}}^{\partial=2}_{\gamma Z,1} \right]
\\
{\cal T}^{p=6}_{62} &=& \frac{g^4}{4 }\left[
2{\cal{Q}}^{\partial=2}_{WW,2}+ 2{\cal{Q}}^{\partial=2}_{WW,3}+ 
2 c_w^2{\cal{Q}}^{\partial=2}_{WZ,7}
+{2} s_w c_w {\cal{Q}}^{\partial=2}_{\gamma ZW,2} + 2 s_w^2 {\cal{Q}}^{\partial=2}_{\gamma W,2}
\right. \nonumber \\ & & \left.
+ \frac{2}{c_w^2} {\cal{Q}}^{\partial=2}_{WZ,2}+ {\cal{Q}}^{\partial=2}_{ZZ,2}
+ \frac{2 s_w}{c_w} {\cal{Q}}^{\partial=2}_{\gamma ZZ,2}
+ \frac{s^2_w}{c^2_w} {\cal{Q}}^{\partial=2}_{\gamma Z,2} \right]
\\
{\cal T}^{p=6}_{63} &=& \frac{g^4}{4 }\left[
{\cal{Q}}^{\partial=2}_{WW,15}+ {\cal{Q}}^{\partial=2}_{WW,16}+ 
\frac{1}{c_w^2}{\cal{Q}}^{\partial=2}_{WZ,22}+{\cal{Q}}^{\partial=2}_{WZ,29}
+ \frac{1}{c_w^2} {\cal{Q}}^{\partial=2}_{ZZ,7}
+ \frac{s_w}{c_w} {\cal{Q}}^{\partial=2}_{\gamma ZW,6}
+ \frac{s_w}{c^3_w} {\cal{Q}}^{\partial=2}_{\gamma ZZ,3} \right]
\\
%
{\cal T}^{p=6}_{64} &=& \frac{g^4}{2 c_w^2 }\left[
2{\cal{Q}}^{\partial=2}_{WZ,8}+ {\frac{1}{c_w^2}}
{\cal{Q}}^{\partial=2}_{ZZ,3} \right] 
\\
{\cal T}^{p=6}_{65} &=& \frac{g^4}{2 c_w^2 }\left[
2{\cal{Q}}^{\partial=2}_{WZ,9}+ {\frac{1}{c_w^2}}
{\cal{Q}}^{\partial=2}_{ZZ,4} \right] 
\\
{\cal T}^{p=6}_{66} &=& \frac{g^4}{2 c_w^2 }\left[
2{\cal{Q}}^{\partial=2}_{WZ,10}+ {\frac{1}{c_w^2}}
{\cal{Q}}^{\partial=2}_{ZZ,5} \right] 
\\
{\cal T}^{p=6}_{67} &=& \frac{g^4}{2 c_w^2 }\left[
 {\cal{Q}}^{\partial=2}_{WZ,11}+ {\frac{1}{c_w^2}}
{\cal{Q}}^{\partial=2}_{ZZ,6} \right] 
\\
{\cal T}^{p=6}_{68} &=& \frac{g^4}{2 c_w^2 }\left[
2{\cal{Q}}^{\partial=2}_{WZ,1}+ c_w^2{\cal{Q}}^{\partial=2}_{ZZ,1} +
2 s_w c_w {\cal{Q}}^{\partial=2}_{\gamma ZZ,1}  +s_w^2{\cal{Q}}^{\partial=2}_{\gamma Z,1} 
\right] 
\\
{\cal T}^{p=6}_{69} &=& \frac{g^4}{2 c_w^2 }\left[
2{\cal{Q}}^{\partial=2}_{WZ,2}+ c_w^2{\cal{Q}}^{\partial=2}_{ZZ,2} +
2 s_w c_w {\cal{Q}}^{\partial=2}_{\gamma ZZ,2} + s_w^2{\cal{Q}}^{\partial=2}_{\gamma Z,2} 
\right] 
\\
{\cal T}^{p=6}_{70} &=& \frac{g^4}{2 c_w^2 }\left[
 {\cal{Q}}^{\partial=2}_{WZ,22}+  {\cal{Q}}^{\partial=2}_{ZZ,7}
+\frac{s_w}{c_w}  {\cal{Q}}^{\partial=2}_{\gamma ZZ,3} 
\right] 
\end{eqnarray}


\end{document}